\documentclass{aa}  

\usepackage{graphicx}
\usepackage[dvipsnames]{xcolor}
\usepackage{txfonts}
\usepackage[colorlinks=true,
    linkcolor=blue,
    citecolor=blue,
    filecolor=magenta,      
    urlcolor=cyan]{hyperref}
\begin{document} 

   \title{Quasi-periodic eruptions from impacts between the secondary and a rigidly precessing accretion disc in an extreme mass-ratio inspiral system}
   \titlerunning{QPEs from EMRI-disc crossing}
   \authorrunning{A. Franchini, M. Bonetti et al.}

   \author{Alessia Franchini
          \inst{1,}
          \inst{2}\fnmsep\thanks{alessia.franchini@unimib.it}
          Matteo Bonetti
          \inst{1,}
          \inst{2,}
          \inst{3}\fnmsep\thanks{matteo.bonetti@unimib.it; second author with equal contribution}
          \and
          Alessandro Lupi
          \inst{4,1,2}
          \and
          Giovanni Miniutti
          \inst{5}
          \and 
          Elisa Bortolas
          \inst{1,}
          \inst{2}
          \and
          Margherita Giustini
          \inst{5}
          \and
          Massimo Dotti
          \inst{1,}
          \inst{2}
          \and
          Alberto Sesana
          \inst{1,}
          \inst{2}
          \and
          Riccardo Arcodia
          \inst{6,7}
          \thanks{Einstein Fellow}
          \and 
          Taeho Ryu
          \inst{8}
    }

   \institute{
            Dipartimento di Fisica ``G. Occhialini'', Universit\`a degli Studi di Milano-Bicocca, Piazza della Scienza 3, I-20126 Milano, Italy
        \and
            INFN, Sezione di Milano-Bicocca, Piazza della Scienza 3, I-20126 Milano, Italy
        \and
            INAF - Osservatorio Astronomico di Brera, via Brera 20, I-20121 Milano, Italy
        \and
            DiSAT, Universit\`a degli Studi dell'Insubria, via Valleggio 11, I-22100 Como, Italy
        \and
            Centro de Astrobiología (CAB), CSIC-INTA, Camino Bajo del Castillo s/n, ESAC campus, Villanueva de la Cañada, E-28692 Madrid, Spain
        \and
            MIT Kavli Institute for Astrophysics and Space Research, 70 Vassar Street, Cambridge, MA 02139, USA
        \and
            Max-Planck-Institut f\"ur extraterrestrische Physik (MPE), Giessenbachstrasse 1, 85748 Garching, Germany
        \and
            Max Planck Institute for Astrophysics, Karl-Schwarzschild-Strasse 1, 85748 Garching, Germany
    }

   \date{Received 31 March 2023 / Accepted 23 May 2023}

 
\abstract{
X-ray quasi-periodic eruptions (QPEs) represent a recently discovered example of extreme X-ray variability associated with supermassive black holes. These are high-amplitude bursts recurring every few hours that are detected in the soft X-ray band from the nuclei of nearby galaxies whose optical spectra lack the broad emission lines typically observed in unobscured active galaxies. 
The physical origin of this new X-ray variability phenomenon is still unknown and several theoretical models have been presented. However, no attempt has been made so far to account for the varying QPE recurrence time and luminosity in individual sources, nor for the diversity of the QPE phenomenology in the different known erupters. We present a semi-analytical model based on an extreme mass-ratio inspiral (EMRI) system where the secondary intersects, along its orbit, a rigidly precessing accretion disc surrounding the primary. We assume that QPEs result from emission from an adiabatically expanding, initially optically thick gas cloud expelled from the disc plane at each impact. 
We produced synthetic X-ray light curves, which we then compared with X-ray data from four QPE sources: GSN 069, eRO-QPE1, eRO-QPE2, and RX J1301.9+2747. Our model aptly reproduces the diversity of QPE properties between the considered objects and it is also able to naturally account  for the varying QPE amplitudes and recurrence times in individual sources. Future implementations will enable us to refine the match with the data and to estimate  the system parameters precisely, making additional use of multi-epoch QPE data. We briefly discuss the nature of the secondary object, as well as the possible implications of our findings for the EMRI population at large.
}

\keywords{galaxies:active -- galaxies:nuclei -- quasars:supermassive black holes -- X-rays: bursts -- Black hole physics -- Relativistic processes}

\maketitle
%
\section{Introduction}

X-ray quasi-periodic eruptions (QPEs) are an extreme and rather puzzling X-ray variability phenomenon that has  recently been detected from the nuclei of nearby galaxies. They are fast bursts in the soft X-ray band, repeating every few hours, superimposed to an otherwise stable quiescent X-ray level that is consistent with emission from a radiatively efficient accretion flow around relatively low-mass massive black holes (MBHs). During these bursts, the X-ray count rate increases by up to two orders of magnitude, depending on the considered energy band. QPEs have thermal-like X-ray spectra with temperature evolving from $kT\simeq 50$-$80$~eV to $\simeq 100$-$250$~eV (and back) in about one to a few hours, with a typical duty cycle of $10$-$30$\%, depending on the specific source. The peak X-ray luminosity is $\simeq 10^{42-43}$~erg~s$^{-1}$. Following their discovery in GSN~069 \citep{Miniutti2019}, QPEs have been observed so far from the nuclei of other four galaxies: RX~J1301.9+2747 \citep{Giustini2020}, eRO-QPE1, and eRO-QPE2 \citep{Arcodia2021}, as well as XMMSL1~J024916.6-04124 \citep{Chakraborty2021}. Only $1.5$ QPEs were detected in  the latter source, but all observed properties indicate it is nevertheless a robust QPE candidate.

We may classify QPE host galaxies as relatively low-mass post-starburst galaxies \citep{Arcodia2021,Wevers2022}, a population that is very similar to that of the hosts of the preferred tidal disruption events \citep[TDEs][]{2020SSRv..216...32F}, with which they also share a preference towards low-mass MBHs of $10^{5-6.7}$~M$_{\odot}$ \citep{Wevers2022}. Two out of the five QPE sources known to date have been associated with X-ray detected TDEs \citep{2023A&A...670A..93M,Chakraborty2021}, and a  new QPE candidate is associated with the X-ray decay following an optically detected TDE \citep{Quintin2023}, further strengthening the QPE-TDE connection. Despite very little X-ray obscuration, none of the QPE galaxies show the typical optical or UV broad emission lines associated with unobscured, type 1 active galactic nuclei (AGNs), although the observed narrow-line ratios suggest the presence of an ionising AGN-like continuum \citep{Wevers2022}. The lack of broad emission lines may be due to an AGN that has switched off sometime in the past leaving relic narrow emission lines or be perhaps associated with TDE-like accretion discs that are likely too compact to support a mature AGN-like broad line region.

Several theoretical models have been proposed to explain QPEs, based on modified disc instabilities \citep{Raj2021,Pan2022,Kaur2022}, gravitational lensing \citep{2021MNRAS.503.1703I}, mass transfer from one or more bodies in different configurations \citep{2020MNRAS.493L.120K,2022MNRAS.515.4344K,2022ApJ...930..122C,2022ApJ...937L..12M,2022A&A...661A..55Z,2022ApJ...941...24K,2022arXiv221008023L,2022ApJ...933..225W,2023ApJ...945...86L}, and collisions between an orbiting secondary object and the accretion accretion flow that is formed around the primary MBH \citep{2021ApJ...917...43S,2021ApJ...921L..32X,2023arXiv230316231L,2023arXiv230403670T}.\footnote{See also \citet{2023MNRAS.521.6143V} and references therein for a crossing model applied to the MBH binary candidate OJ287.}  
Although these models can explain some of the features observed in the light curves of the confirmed QPE sources (mainly GSN~069) there is no comprehensive model able to reproduce the variety of behaviours observed in the QPEs of different sources. Here, our efforts are aimed at building a comprehensive and flexible model that can reproduce these variabilities and provide a possible emission mechanism that is compatible with the observations.

We propose a new model in which QPEs are produced by extreme mass ratio inspirals (EMRIs). Those are binaries in which a lighter companion (usually a stellar mass black hole, BH) adiabatically inspirals onto a MBH under the effect of gravitational wave (GW) back-reaction. In the specific scenario we propose, the observed X-ray flares emerge as the companion crosses an accretion disc that is rigidly precessing around the central MBH. This is somewhat similar to the model proposed by \cite{2021ApJ...921L..32X}, but it also includes the relativistic Lense-Thirring \citep{Lense1918} precession of the accretion disc.
We include both apsidal and nodal precession of the EMRI companion up to 3.5PN order beyond Newtonian dynamics. Our proposed model has the flexibility to reproduce the wide variety of QPE light curves including luminosity variations and timing irregularities, owing to the various orbital and precession frequencies involved in the aforementioned processes and their interplay. 
We emphasise that our aim, at this stage, is not to apply our model to a specific source for parameter inference, but to show that the different amplitudes and timing properties of the QPE phenomenon can be qualitatively reproduced.

The paper is organised as follows. In Section \ref{sec:methods}, we describe our model in detail  and we present the resulting light curves in Section \ref{sec:results}. We then draw our conclusions and discuss future improvements of the model in Section \ref{sec:concl}.

\section{Methods}
\label{sec:methods}

\begin{figure}
\centering    \includegraphics[width=0.7\columnwidth]{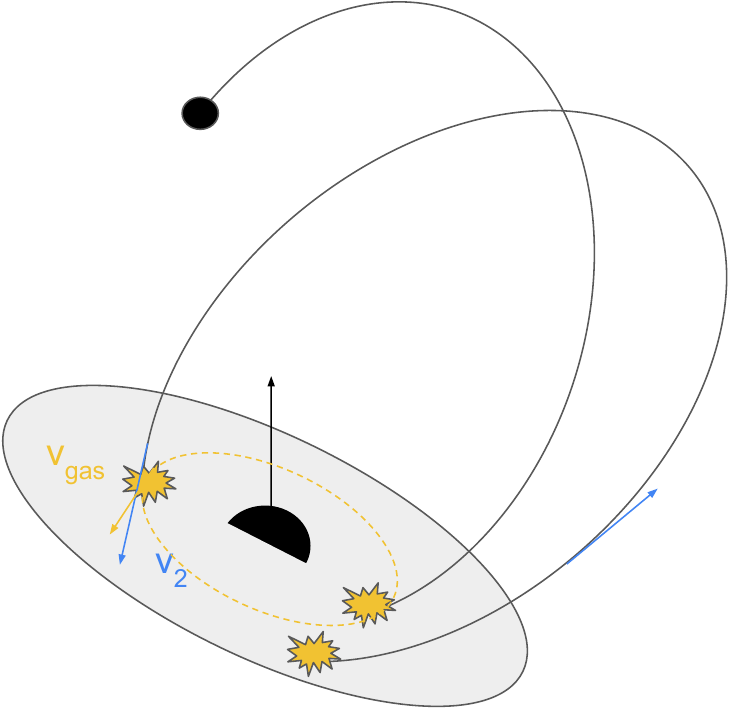}
    \caption{Schematic view of the model where the central big black dot represents the primary MBH, the small black dot represents the EMRI companion, the grey area represents the tilted disc and the EMRI-disc crossings are shown with yellow stars. The vertical arrow shows the MBH spin vector. The misalignment between the EMRI orbit and the disc is enhanced in this cartoon with the purpose of showing the impacts more clearly. However, we use smaller misalignments to produce the QPEs light curves. The blue and yellow arrows show the EMRI and gas velocity vectors, respectively, while the yellow dashed line only mimics an orbit inside the disc.}
    \label{fig:sketch}
\end{figure} 

Using a semi-analytic approach, we modelled a system composed of a MBH with mass, $M_1$, and spin parameter, $\chi$, surrounded by a misaligned and rigidly precessing accretion disc, whose properties are compatible with a disc generated by a stellar TDE (see Section~\ref{sec:disc}). The MBH is further orbited by a stellar-mass BH (the EMRI companion, hereafter) with a mass, $M_2$, that intersects the accretion disc (up to three times per orbit; see Section~\ref{sec:crossing}). The mass ratio of the system $q = M_2/M_1$ is assumed to be $q\ll 1$ throughout the paper -- a feature that allows us to primarily characterise the binary system as an EMRI.

\begin{table}
    \centering
    \begin{tabular}{p{3.8cm}|p{2cm}|p{1cm}}
    \hline
        Quantity & Range & Units \\
        \hline
        \hline
         primary spin \,\, $\chi$ & 0.1 -- 0.65 & - \\
         disc inclination \,\, $i_{\rm disc}$ & $5^{\circ}$ -- $30^{\circ}$ & - \\
         EMRI inclination \,\, $i_{\rm EMRI}$ & $6^{\circ}$ -- $20^{\circ}$ & - \\
         EMRI eccentricity \,\, $e$ & 0.05 -- 0.5 & - \\
         disc mass \,\,$M_{\rm d}$ & 0.01 -- 4 & $M_{\odot}$\\
         EMRI semi-major axis \,\,$a$ & 50 -- 355 & $R_{\rm g}$ \\
         primary mass \,\,$M_1 $ & $10^{5}$ -- $10^{6.3}$ & $M_{\odot}$ \\   
         secondary mass \,\, $M_2$ & 100 & $M_{\odot}$ \\
         \hline
    \end{tabular}
    \caption{Parameters of the model that we adjust within these ranges to reproduce the different behaviours of QPE sources. }
    \label{tab:params}
\end{table}

\subsection{EMRI post-Newtonian trajectory}

We integrated the equations of motion of the EMRI system considering post-Newtonian (PN) dynamics and retaining corrections up to 3.5 PN order. Integer terms such as 1, 2, or 3 PN represent conservative terms, while 2.5 and 3.5 PN orders are dissipative terms accounting for GW emission. Finally, the 1.5 (conservative) PN term takes into account the effect of the MBH spin, specifically: the leading order spin-orbit coupling. 
We consider the equations of motion derived using a Lagrangian formalism and referred to the relative position and velocities of the binary components. Practically,
we evolve forward in time the relative separation, $\mathbf{r,}$ and velocity, $\mathbf{v,}$ of the EMRI components according to: 
\begin{equation}
    \label{eq:PN_acc}
    \frac{d^2 \mathbf{r}}{dt^2} = -\frac{G M}{r^2} \biggl( (1+\mathcal{A}) \ \mathbf{n} + \mathcal{B} \mathbf{v} \biggr) + \mathbf{C}_{1.5} + \mathcal{O}\left(\frac{1}{c^8}\right), 
\end{equation}
where $\mathbf{n} = \mathbf{r}/|\mathbf{r}|$, while $M$ is the binary total mass. The coefficients $\mathcal{A}$ and $\mathcal{B}$ can be found in \citet{Blanchet2014} and encode the interaction due to the non-spin terms, while the coefficient $\mathbf{C}_{1.5}$ introduces the leading order spin-orbit interaction and can be retrieved from \citet{2013CQGra..30g5017B}.
We employed a \textsc{c++} implementation of the Bulirsch-Stoer algorithm to integrate the above differential equations. More details can be found in \cite{2016MNRAS.461.4419B}, except that here we do not include the interaction with a third body or the hardening of the binary due to the presence of a stellar cluster, with the latter assumption being justified at such small separations.
The main advantage of this approach, compared to the one adopted by \citet{2021ApJ...921L..32X}, is that we are not limited by the test particle approximation and therefore in the choice of the EMRI companion mass.

\subsection{Disc model}
\label{sec:disc}
 
Based on the growing observational evidence that QPEs arise after a TDE-like precursor \citep{Chakraborty2021,2023A&A...670A..93M}, we considered a disc formed by a tidal disruption of a star by the primary MBH and modelled it following \cite{Franchini2016}.
The main difference between the current model and \cite{Franchini2016} concerns the accretion rate onto the central MBH, that in the current investigation is set to reproduce the observed nuclear activity when QPEs are not occurring (the "quiescence" luminosity, hereafter).

More in detail, the disc extends all the way to the innermost stable circular orbit (ISCO), which is a function of the primary BH spin $\chi$ \citep{1972ApJ...178..347B}, and we assumed the disc to be prograde with respect to the spin of the MBH, i.e. the $z$ component of its angular momentum is positive. 

We assumed the mass of the disc to be $M_{\rm d}=f\,M_{\odot}$ with $f\leq 4$ and to be distributed with a power law surface density profile $\Sigma = \Sigma_0(R/R_{\rm g})^{-p} $ out to a radius $R_{\rm out}= 200-400 R_{\rm g}$, depending on the primary mass, where $R_g ={G M_1}/{c^2}$ is the MBH gravitational radius. The normalization of the density profile can be therefore computed in a straightforward way as: 

\begin{equation}
    \Sigma_0 = \frac{M_d(2-p)}{2\pi R_{\rm g}^2}\left(\left(\frac{R_{\rm out}}{R_{\rm g}}\right)^{2-p}-\left(\frac{R_{\rm ISCO}}{R_{\rm g}}\right)^{2-p}\right)\,,
\end{equation}
with $p$ as the power law index.

We define the disc aspect ratio as \citep{Strubbe2009,Franchini2016} 
\begin{equation}
    \frac{H}{R} = \frac{3}{2}(2\pi)^{1/2}\eta^{-1} K(R)^{-1}\dot{m}_1\left(\frac{R}{R_{\rm g}}\right)^{-1}
,\end{equation}
where
\begin{equation}
     K (R) = \frac{1}{2}+\left[\frac{1}{4}+6 \left(\frac{\dot{m}_1}{\eta}\right)^{2}\left(\frac{R}{R_{\rm g}}\right)^{-2}\right]^{1/2} \,.
\end{equation}
The $K(R)$ coefficient takes into account the modified structure for a slim disc. Here, $\eta=0.1$ is the accretion efficiency and $\dot{m}_1 = \dot{M_1}/\dot{M}_{1,\rm Edd}$, namely, the mass accretion rate of the central MBH normalised to its Eddington limit. 

We assumed that the angular momentum of the disc is misaligned with respect to the MBH spin and that the warp induced by the Lense-Thirring (LT) effect \citep{Lense1918} propagates as a bending wave allowing the disc to rigidly precess around the MBH \citep{Franchini2016}. We modelled the disc viscosity according to the Shakura \& Sunyaev $\alpha$-prescription \citep{SS1973}. In the assumed bending waves regime, $\alpha$ is lower than $H/R,$ so we can reasonably assume that the alignment timescale is long enough to neglect any disc-spin alignment effects  \citep{SL2012,Franchini2016}. 

Under these assumptions,\footnote{We assume the misalignment between the disc angular momentum and the MBH spin to be small, so that we can apply linear theory for the warp propagation inside the disc. } the disc is predicted to precess rigidly with a frequency $\Omega_p$ that is the angular-momentum-weighted average of the Lense-Thirring precession $\Omega_{\rm LT}(R)$ frequency over the extent of the disc:

\begin{equation}
    \Omega_p = \frac{\int_{R_{\rm ISCO}}^{R_{\rm out}} \Omega_{\rm LT}(R) L(R) 2\pi R dR }{\int_{R_{\rm ISCO}}^{R_{\rm out}} L(R) 2\pi R dR },
\end{equation}
with the Lense-Thirring frequency given by:
\begin{equation}\label{eq:lt}
    \Omega_{\rm LT}(R) = \frac{c^3}{2 G M_1}\frac{4\chi\left(\frac{R}{R_g}\right)^{-3/2} - 3\chi^2\left(\frac{R}{R_g}\right)^{-2}}{\left(\frac{R}{R_g}\right)^{3/2} + \chi},
\end{equation}
and with $L(R) = \Sigma \Omega(R) R^2$ being the angular momentum modulus at a radius, $R$. We assumed that the gas orbits the MBH on circular Keplerian orbitsl thus, the gas orbital frequency and its velocity modulus are simply given by $\Omega(R) = (G M_1/R^3)^{1/2}$ and $|\mathbf{v}_{\rm gas}(R)| = (G M_1/R)^{1/2}$, respectively.

We stress that we have assumed a disc model broadly consistent with the tidal disruption of a star by the primary MBH. However, the disc properties are poorly constrained parameters and we can therefore vary them to reproduce different observed timing behaviours in QPE sources.

We assumed that the disc is precessing rigidly around the primary MBH, as there has been (to date) no precise criterion for a disc to precess rigidly other than order of magnitude estimates that compare the sound crossing time to the viscous time, with the latter being largely unconstrained. If the annulus of the disc that corresponds to the crossing does precess with its own frequency, then this will probably cause a different modulation in the light curves. We leave the investigation of a differentially precessing disc annuli to a future work.

We further note that precession is not crucial when comparing our model with single-epoch QPE light curves, as it induces a modulation on timescales that are longer than the typical observation duration ($\sim 1.5$~days, except in the case of eRO-QPE1). Future, multi-epoch observations of QPE sources are needed to detect unambiguously the putative modulations induced by disc precession.

\subsection{Disc crossing}
\label{sec:crossing}

The EMRI companion crosses the disc at different radii between one and three times per orbit around the MBH since the encounter can occur at pericentre and along the precessing disc line of nodes (similar to the model of non periodic outburst in Be/X-ray binaries proposed by \citealt{Martin2021}).

The apsidal and nodal precession of the EMRI companion further complicate the time series of disc crossings making the interval between subsequent episodes change during time, as similarly observed during the different campaigns in the light curve of RX J1301.9+2747 \citep[][Giustini et al., in prep.]{Giustini2020}.

In our model, the EMRI evolution is computed from the numerical integration of Eq.~\ref{eq:PN_acc}, thus the characterisation of the disc crossings has to be done numerically as well. We stress that in our model, the disc is represented by a rigidly rotating plane that does not influence the EMRI evolution. In order to completely characterise the crossings and infer the numerical values of the physical quantities involved, we proceed as follows. 
As a first step, we identify the times and positions at which the EMRI companion crosses the disc plane.
Specifically, those are found along the EMRI trajectory when the coordinates $\mathbf{r}_2 = (x_2,y_2,z_2)$ of $M_2$ satisfy the relation $w_x x_2 + w_y y_2 + w_z z_2 = 0$. Here, $\mathbf{w} = (w_x,w_y,w_z)$ represents the vector normal to the disc plane, which changes with time according to\footnote{We assume that the direction of the MBH spin lies along the $z$-axis.}
\begin{align}
    w_x &= \sin(\Omega_{\rm p} t) \sin(i_{\rm disc}),\\
    w_y &= \cos(\Omega_{\rm p} t) \sin(i_{\rm disc}),\\
    w_z &= \cos(i_{\rm disc}).
\end{align}
Given the crossing location, we can determine the velocity vector of the gas ($\mathbf{v}_{\rm gas}$) at each passage. We calculate the components of $\mathbf{r}_2$ in the disc plane and then the polar angle $\phi_{\rm d} = \arctan(y_{2,{\rm d}}/x_{2,{\rm d}})$ in such a frame. Since the gas is assumed to be in circular motion, we know that the the gas velocity has to be perpendicular to the projection of $\mathbf{r}_2$ in the disc plane $\mathbf{r}_{2,{\rm d}}$, thus the gas velocity in the reference plane of the disc is given by:
\begin{equation}
    \mathbf{v}_{\rm gas,{\rm d}} = \left(\frac{G M_1}{R_{\rm cross}}\right)^{1/2} ( -\sin(\phi_{\rm d}), \cos(\phi_{\rm d}), 0 ),
\end{equation}
where $R_{\rm cross}$ is the radius at which the crossing occurs.
Finally, the  velocity vector above is rotated back in the initial frame and the relative velocity between the gas and $M_2$ is evaluated from the vector difference, $\mathbf{v}_{\rm rel} = \mathbf{v}_{\rm gas} - \mathbf{v}_2$.
We assumed the relative velocity does not change during the crossing as this occurs on a short timescale. For instance, for GSN 069, this is $ 2H/v_2 \sim 200$ s.

We note that the nodal LT precession of the disc does in principle make it possible for the precession velocity to eventually point in the opposite direction with respect to the EMRI velocity. However, in all of our cases, the precession of the disc is  slower than the orbital period and the precession velocity  always points in the same direction as the orbital one. In other words, we never have a full precession of the disc within an EMRI orbital period.

The crossing times determined by the above procedure are referred to the reference frame of the binary. In order to relate the time at which the crossing occurs to the arrival time of emitted photons at the observer, we included the relevant delays, specifically: the Roemer, Shapiro, and Einstein delays. These delays take into account the fact that, once photons are released by the EMRI companion crossing the disc, they have to propagate towards a distant observer, thus they are influenced by the light-travel time within the binary system (Roemer delay $\Delta_R(\tau)$) and the non-zero curvature generated by the primary companion which bends the travel path (Shapiro $\Delta_S(\tau)$ and Einstein delays $\Delta_E(\tau)$).
Specifically, we express the arrival time to the distant observer as \citep[see e.g.][sec. 10.3.6]{2014grav.book.....P} as:
\begin{equation}
    t_a = \tau + \Delta_R(\tau) + \Delta_S(\tau) + \Delta_E(\tau),
\end{equation}
where the three delays are given by the following expression:
\begin{align}
    \Delta_R(\tau) &= |\mathbf{r}_{\rm obs}-\mathbf{r}_{2}|/c,\\
    \Delta_S(\tau) &= \frac{2GM_1}{c^3}\ln\left(\frac{|\mathbf{r}_{\rm obs}-\mathbf{r}_{1}| + (\mathbf{r}_{\rm obs}-\mathbf{r}_{1})\cdot \mathbf{k}}{r + \mathbf{r}\cdot\mathbf{k}}\right),\\
    \Delta_E(\tau) &= \frac{M_2+2M_1}{M}\sqrt{\frac{a^3}{G M}}\frac{G M_1}{a c^2} \ \sin(u).
\end{align}
In the above equations, $M = M_1 + M_2 $ is the mass of the binary, $\mathbf{r}_{\rm obs}$ denotes the observer position, $\mathbf{r}_{1}$ and $\mathbf{r}_{2}$ represent the position vectors of the primary and secondary BH with respect to the centre of mass of the system, while the unit vector $\mathbf{k}$ is given by:
\begin{equation}
    \mathbf{k} = \frac{\mathbf{r}_{\rm obs}-\mathbf{r}_{2}}{|\mathbf{r}_{\rm obs}-\mathbf{r}_{2}|}.
\end{equation}
Here $a$, $e$, and $u$ respectively represent the orbit semi-major axis, eccentricity and eccentric anomaly, respectively.

\subsection{Emission mechanism}
\label{sec:emission}

Since we are interested in modelling the light curve that results from the disc crossings, we need to investigate possible emission mechanisms that can reproduce the amplitude, duration, and spectral properties of QPEs.

The emission mechanisms that have thus far been proposed in the literature to explain QPEs, within the disc-crossing model, consider either accretion onto the secondary object or the emission generated by the shock as the companion passes through the accretion flow \citep{2021ApJ...917...43S,2021ApJ...921L..32X}. We can safely rule out the first hypothesis since, in order to produce the observed luminosities of $\sim 10^{42}-10^{43}$ erg s$^{-1}$, the accretion onto the EMRI companion would greatly exceed its Eddington limit.
As for the shock mechanism, this has been investigated and modelled extensively in \cite{2004A&A...413..173N} where the authors considered a star that passes through an accretion disc surrounding a MBH, generating a shock due to the compression of the disc material in front of and aside of it. However, some of the assumptions in their work are tailored to the specific case of Sgr~A$^*$ and are not well suited to the case of the radiatively efficient, high-mass accretion rate discs that are relevant for QPE sources.

Here, we instead consider the emission from a gas cloud that is pulled out of the disc as a consequence of the crossing of the EMRI companion, broadly following the approach outlined in \citet{2016MNRAS.457.1145P}. We assume the gas cloud to have an initial radius, $R_{\rm in}$, that is of the order of the influence radius of the the secondary object that crosses the disc, namely:
\begin{equation}
    R_{\rm in} \sim R_{\rm inf} = \frac{G M_2}{c_{\rm s}^2+v_{\rm rel}^2},
\end{equation}
where $c_{\rm s}$ is the disc sound speed, while $v_{\rm rel}$ is the modulus of the relative velocity (see previous section for details). We consider the cloud to be optically thick (we assume its density to be equal to the disc density at the crossing radius) and with a post-shock temperature of $\sim10^6$ K, which corresponds to the observed QPE emission of $\sim 100-120$ eV.
The exact value of the cloud initial (post shock) temperature, $T_2$, depends on the source considered and it is determined by using the Rankine-Hougoniot condition for shocks in a radiative pressure dominated gas disc:
\begin{equation}
    T_2 = \left[1+\frac{8}{7}\left(\mathcal{M}_{\rm e,1}^2-1\right)\right]^{1/4} T_1
,\end{equation}
where $T_1$ is the pre-shock disc gas temperature and $\mathcal{M}_{\rm e,1}$ is the effective pre-shock Mach number, given by:
\begin{equation}
    \mathcal{M}_{\rm e,1} = \left(3\gamma \frac{ \mathcal{M}_1^2}{4R_{\rm p}}\right)^{1/2}
,\end{equation}
where $\mathcal{M}_1 = v_{\rm rel}/c_s$ is the pre-shock Mach number, $R_{\rm p}$ is the ratio between the radiation and gas pressure before the shock occurs, and $\gamma=4/3$ is the adiabatic index. The pre-shock temperature $T_1$ is determined by the properties of the disc and is obtained from the relation linking the total pressure to the gas sound speed, 
namely:\footnote{Eq.~\ref{eq:pressure} can be analytically solved for $T_1$ but being the expression rather lengthy we avoid to write it explicitly. The equation can also be easily solved numerically.}
\begin{equation}
    \label{eq:pressure}
    P = P_{\rm rad} + P_{\rm gas} = \frac{\rho c_s^2}{\gamma} ,
\end{equation}
with the radiation and gas pressure respectively given by
\begin{align}
    P_{\rm rad} &= \frac{4}{3}\frac{\sigma_{\rm sb}T_1^4}{c},\\
    P_{\rm gas} & = \frac{\rho k_{\rm B} T_1}{m_{\rm p}\,\mu},
\end{align}
with $\sigma_{\rm sb}$ being the Stefan-Boltzmann constant, $k_{\rm B}$ the Boltzmann constant, $\mu=1$ the mean molecular weight, and $m_{\rm p}$ the proton mass.

We note that our model implies that $T_2/T_1$ is a decreasing function of mass accretion rate $\dot{m}_1$, asymptotically reaching $T_2/T_1$ at high $\dot{m}_1$. As QPEs only carry a small fraction of the disc bolometric luminosity \citep{2023A&A...670A..93M}, they become undetectable once $T_2/T_1 \simeq 1$. Our result appears therefore to be consistent with observations of GSN~069 where QPEs are not detected above a disc luminosity threshold that corresponds to an Eddington ratio of $\simeq 0.4\pm 0.2$ \citep{2023A&A...674L...1M}.

We assumed the temperature of the cloud to increase from $T_1$ to $T_2$ instantaneously, giving us a fast rise in the QPE profile. 
This is an assumption we made for simplicity, as we are mostly interested in reproducing the timing and peak luminosity properties rather than the exact individual QPE profile (see Sect.~\ref{sec:results}). This cloud of gas might take some time to exit from the disc and the gas would also have to thermalise before emitting at $T_2$. This assumption can likely be relaxed by considering the effects of photon diffusion and disc crossing timescales \footnote{We have calculated the timescale for the cloud to cross the disc as $2H/v_{\rm rel}$, using GSN 069 as test case, finding this to be $\sim360$ s. 
This can explain the rise in the luminosity peak observed in the QPE profiles. The thermalisation of the photons in the cloud is instead almost instantaneous. 
The photons diffusion timescale is, again, for GSN 069, on the order of $\sim 10^6$ s. We defer the investigation of these effect to future work.} that induce a rise on a slightly shorter timescale compared to the decay. 

We assumed the gas cloud to emit black body radiation while it adiabatically expands after leaving the dense and pressure confining accretion disc environment. The actual properties of the hot atmosphere around the accretion disc are not observationally constrained and not derivable from first principles in our model. However, QPEs can only be detected if their peak temperature, $T_2$, is significantly higher than the pre-shock disc emission, $T_1$. Observationally, QPEs are characterised by $T_2\simeq 2-3\times T_1$. We therefore tailored the expansion of the ejected cloud so that the initial temperature drops by a factor of $2-3$ during the typical QPE duration assuming adiabatic expansion.
The radius of the cloud changes over time as:
\begin{equation}
    R(t) = R_{\rm in}+ \frac{2R_{\rm in}}{\Delta t_{\rm QPE}} t
     \label{eq:Rdef}
,\end{equation}
where $\Delta t_{\rm QPE}$ is the duration of the QPE starting from the time of the impact, while the temperature at which the black body spectrum is emitted changes with time as $T_{\rm exp} = T_2 (R_{\rm in}/R(t))$\footnote{We note that the radiative cooling of the cloud is negligible on the expansion timescale. }. 

Assuming the cloud initial radius to be $R_{\rm in} \sim 10^{11}$~cm and for it to expand by a factor of 2-3 in $\sim 1$ hr, we infer an expansion velocity $v_{\rm exp}\sim 400$~km~s$^{-1}$. This is consistent with the estimate $v_{\rm exp}\sim v_{\rm rel}\left( R_{\rm in}/H\right)$ \citep{1980ApJ...237..541A,2016MNRAS.457.1145P}, with $v_{\rm rel}$ being the relative velocity between the EMRI companion and the disc, which we infer in the  case of GSN 069 from our model.
We note that the assumed expansion factor is consistent with the observed data of GSN~069  (e.g. lower panel of Figure 18 in \citealt{2023A&A...670A..93M}).

The emitted luminosity in the soft-X ray band $0.2-2$ keV is therefore:
\begin{equation}
    L_{\rm X} = 4\pi R(t)^2 \int_{0.2{\rm keV}}^{2{\rm keV}} \frac{2h\nu^3}{c^2}\frac{d\nu}{e^{h\nu/k_{\rm B}T_{\rm exp}}-1}\,.
    \label{eq:luminosity}
\end{equation}

As discussed in Sect.~\ref{sec:secondary}, we specifically consider the case in which the secondary is a BH of mass, $M_2$. We assume the initial cloud radius $R_{\rm in}$ to be either determined by the secondary influence radius, $R_{\rm inf}$, or by the disc half-height at the crossing radius, $H_{\rm cross}$, depending on which one is the smallest. For our choice of parameters we have $R_{\rm inf}\lesssim H_{\rm cross}$ for all the sources we consider. 

We then construct the light curve by assuming that the emission starts at the time of the impact. 
We then assign a time length that corresponds to the observed QPE duration (therefore, different for each source and tailored to the specific observed burst duration for that source) and an amplitude that follows the decay in the luminosity according to Eq. (\ref{eq:luminosity}).

\begin{figure}
    \centering
    \includegraphics[width=\columnwidth]{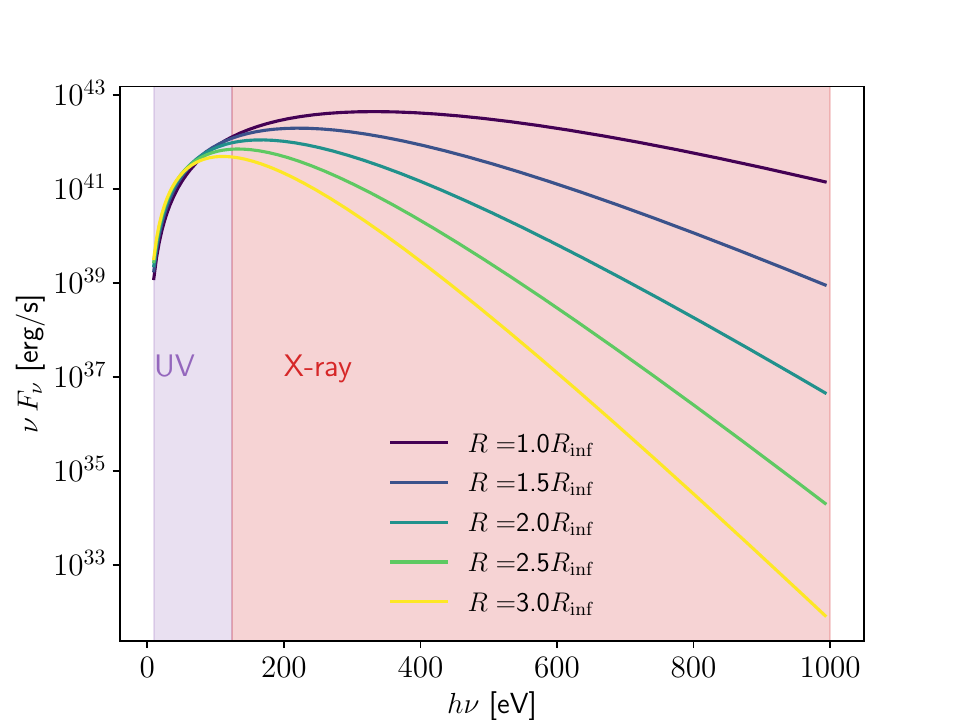}
    \caption{Spectrum emitted by a cloud of initial size $R_{\rm inf}$ and initial temperature $T=10^6$ K that expands by a factor of $3$ in size over one hour.}
    \label{fig:emission}
\end{figure}

As an example, Figure \ref{fig:emission} shows the spectrum of the radiation emitted in the band $10$ eV $-2$ keV by a $10^6$ K cloud that detaches from the disc as a consequence of the EMRI companion passage through it. The cloud initial size is $R_{\rm in}=R_{\rm inf}$ and it is assumed to expand by a factor of $3$ in radius over the 1 hr duration of the observed QPE signal. 
The exact value depends on $R_{\rm inf}$, which in turn depends on the relative velocity between the EMRI companion and the disc -- and, thus, ultimately on the source considered.

\subsection{The secondary object}
\label{sec:secondary}

The QPE timing properties and the initial temperature of the expanding cloud producing them, $T_2$, are set by the disc properties and by the EMRI dynamics. Thus, these parameters are independent of the nature and mass of the secondary, as long as $q\ll 1$. On the other hand, the QPE (peak) luminosity depends on the initial size of the expanding cloud, $R_{\rm in}$, as shown in Eqs.~(\ref{eq:Rdef}) and (\ref{eq:luminosity}). Thus, only impacts that form clouds with $R_{\rm in} \simeq 10^{11}$~cm are able to produce $L_X \simeq 10^{42}-10^{43}$~erg~s$^{-1}$ at QPE peak, as observed. 

For compact companions such as BHs, neutron stars, and white dwarfs, the initial size of the expanding cloud  is $R_{\rm in} \sim R_{\rm inf}$ and it thus depends on companion mass and relative velocity at impact, $v_{\rm rel}$. It is easy to show that BHs with masses $\ll 100$~M$_\odot$ never reach the required $R_{\rm in}$ even for very low $v_{\rm rel}$. The same applies to neutron stars and white dwarfs that can therefore be safely ruled out with respect to the framework of the simple emission mechanism we have considered. Slightly less massive BHs, that is, around and above 30-40 M$_\odot$ are still viable candidates given that the luminosity of the flares is also significantly influenced by the disc properties (e.g. denser clouds produce brighter flares in our model).
On the other hand, the lifetime of the EMRI system, regulated by the GW emission, becomes too short for significantly more massive BHs to be consistent with the recurrence rate of the QPEs. Thus, BHs with mass $\sim 100$~M$_\odot$\footnote{We chose 100 M$_\odot$ as a reference value with the understating that slightly less massive stellar BH cannot be completely ruled out. We plan to explore this point in a more detailed follow-up study.} can potentially produce clouds with sufficiently large $R_{\rm in}$ provided that $v_{\rm rel}$ is low enough; that is to say, if the secondary orbit is prograde with respect to the accretion disc and the misalignment between the orbit and disc planes is not too large. This is precisely the situation we assume in our model.

In general, we cannot completely rule out that the companion responsible for the QPEs is a star, as has been considered in the very recent works by \citet{2023arXiv230316231L} and \citet{2023arXiv230403670T}. A typical main sequence star has indeed a radius comparable to the required $R_{\rm in}$ and, thus, it could  (in principle) produce QPEs of a high enough luminosity with the same mechanism outlined in Section \ref{sec:emission}. However, assuming $R_{\rm in}$ to be on the order of the stellar radius would make the initial cloud size independent of $v_{\rm rel}$, making it difficult to produce the modulation of the luminosity observed in almost every QPEs source. The initial cloud temperature for stars-disc collisions is expected to be significantly lower than the typical QPE one, but \citet{2023arXiv230316231L} have shown that Comptonisation in the highly ionised medium can actually produce soft X-ray emission. 
On the other hand, such a scenario poses two additional constraints to consider. First, the pericentre distance of the secondary, constrained from the observed recurrence of QPEs has to be larger than the tidal radius $R_{\rm t}\approx(M_1/M_\odot)^{1/3}R_\odot$, namely, the radius at which the star is torn apart by the tidal field of the primary MBH. In principle, if QPEs are only produced by impacts, the star also needs to avoid overfilling its Roche lobe at pericentre --  otherwise, any mass transfer will likely contribute to the QPE phenomenon as well. Second, we have to compare the energy of the impact with the star's binding energy to determine whether the star can actually survive the crossing. 

Considering the orbital parameters we obtained (as described in Sect.~\ref{sec:results} below and which are basically independent on the nature and mass of the secondary for $q\ll 1$), we note that the secondary pericentre distance is well within the tidal radius of a main sequence star for RX~J1301.9+2747, and only marginally outside for GSN~069. We further note that in three (GSN 069, eRO-QPE2, and RX J1301.9) out of the four QPE sources considered here, a solar-type star that orbits the primary MBH with the inferred separations would overfill its Roche lobe, therefore transferring mass to the primary. We caution that these estimates depend on the primary MBH masses, which are not well constrained, and on the complex physics driving TDEs and mass transfer; however, we point out that these considerations must be taken into account when discussing the possible nature of the EMRI companion. We have also computed the impact energy for different sources and compared it with a solar mass star binding energy. We find the impact energy to be (at best) comparable to the binding energy. This would suggest that the star is unlikely to survive many passages through the disc without significantly modifying its structure. \cite{2023arXiv230316231L} discarded BH companions based on the influence radius size argument. However, we have noted that the estimate in their Eq. A1 considers the orbital velocity instead of the relative velocity with respect to the disc. This is valid only under the assumption that the companion velocity is much larger than the disc velocity and this does indeed lead to an influence radius smaller by about two orders of magnitude, as compared to the case of prograde orbits with modest misalignments that we study here.

In summary, QPEs induced by impacts between the secondary EMRI component and the accretion disc around the primary MBH can (in principle) be produced either by massive stellar BHs \citep[which we know to exist in virtue of LIGO/Virgo findings;][]{2021arXiv211103606T} in prograde orbits (i.e. with low $v_{\rm rel}$), or by stars. We focus here on the former scenario for the reasons given above and since this avoids complications due to the detailed treatment of the star structure (and its evolution due to the impacts) and the possible onset of mass transfer events. As already mentioned, the timing of QPEs is essentially independent from the nature and mass of the companion, so that the orbital parameters derived in Sect.~\ref{sec:results} can be considered to be general, while the individual QPE luminosity does actually depend on the secondary nature and on the envisaged detailed emission mechanism.

Finally, another relevant aspect to consider  when discussing the nature of the secondary object that triggers the QPE is that when assuming a typical evolved stellar mass function (e.g. \citealt{2001MNRAS.322..231K}), the number density of stellar BHs is $\sim 10^{-3}$ that of normal stars. However, stellar BHs are heavier than normal stars and segregate closer to the MBH owing to dynamical friction. This mass segregation is generally completed in a fraction of the relaxation time and results in the well-known \citet{1977ApJ...216..883B} cusp: stellar black holes distribute into a steep cusp with density $\propto r^{-1.75}$, while lighter stars follow a milder power-law cusp $\propto r^{-1.35}$ \citep{2022MNRAS.511.2885B}. The cusp extends within a radius close to the influence radius of the central MBH.  It has been shown that typical nuclear star clusters hosting the relatively low-mass MBHs involved in QPEs feature short relaxation times (shorter than a Gyr, \citealt{2022MNRAS.514.3270B}); thus, they are very likely to be segregated and dynamically relaxed. More specifically, we can safely assume mass segregation and the \citeauthor{1977ApJ...216..883B} cusp is in place only below $\sim 1$ pc (which is close to the influence radius of the primary MBH). Thus, at a 1 pc scale, we can assume the number density of stellar BHs to be $\sim 10^{-3}$ that of stars and extrapolate their number density at smaller scales, considering that stellar BHs and stars have distributions with different profiles ($r^{-1.75}$ vs $r^{-1.35}$).
It is easy to see that -- at radii of the order of the inferred semi-major axis at which QPEs are occurring (i.e. $\sim 100 R_g\sim 10^{-6}-10^{-5}$ pc as presented in the next section) -- the number density of stellar black holes is comparable to that of stars. This consideration, together with the fact that stars may be tidally destroyed at such small separations or consumed by the repeated interactions with the disc, seems to support the scenario in which a stellar BH is generating the QPEs. 

Another interesting feature of QPEs is that, as mentioned in the introduction, their host galaxies seem to overlap with the typical hosts of tidal disruption events. These galaxies feature a rate of TDEs that is $\sim10^{-4}$ per year or even greater \citep[e.g.][]{2020SSRv..216...32F, 2021ApJ...908L..20H}. As mentioned above, QPEs may involve the remnant disc of a tidal disruption event. Assuming the nuclear stellar cluster of the host galaxy is mass-segregated and dynamically relaxed, the event rates of EMRIs can be shown to be $\sim 10^{-2}-10^{-3}$ times the event rates of tidal disruption events (Bortolas et al., in prep.). Yet EMRIs may take $\sim 10^5$ years to complete their inspiral, as GWs are initially relatively inefficient at inducing their orbital decay, while tidal disruptions happen promptly and remain observable over timescales of years at most. 
A quantitative assessment of this statement will be presented in an upcoming study.

\section{Results}
\label{sec:results}

We used the model developed in Sect.~\ref{sec:methods} to produce the synthetic light curves of four observed QPE sources: GSN 069, eRO-QPE2, eRO-QPE1, and RX J1301.9+2747. The value of the primary mass $M_1$ is crucial to correctly match the QPEs timing properties, as it sets the scale of the system, and we constrained it through the M-$\sigma$ relation \citep{Wevers2022}. The derived $M_1$ is therefore associated with significant uncertainties. The semi-major axis of the EMRI system was then constrained by the observed QPE recurrence period for each source. We further assumed the secondary to be a $M_2 = 100\,M_{\odot}$ Schwarzschild black hole for all sources.  
We note that varying values of $M_2$ changes the radius of influence $R_{\rm inf}$ and, therefore, the relative velocity would need to be adjusted to match the QPE peak luminosities. Changing the relative velocity is likely to lead to slightly different orbits and, consequently, timing properties.
The orbital eccentricity, the MBH spin, and the disc parameters, as well as the misalignment angles of the EMRI orbit and disc with respect to the plane perpendicular to the MBH spin, do not have stringent observational constraints. Therefore, they must be tuned to reproduce the observed light curves. 

Our aim is to qualitatively reproduce the recurrence times and relative X-ray luminosities of QPEs in individual sources and to account for the diversity of QPE properties in the different objects. In principle, our model can be used to infer the system parameters accurately from the observed (often multi-epoch) available light curves and we plan on building a Bayesian inference algorithm to set a meaningful posterior on the model parameters. This will be the subject of a follow-up paper. Here, we do not focus on the individual QPE profile, preferring to defer a detailed study of the involved micro-physics to a future work. For each source, we also provide an estimate of the timescale over which GWs drive the EMRI to merge. 

\subsection{Synthetic light curves}

\subsubsection{GSN 069}
\label{subsection:GSN}

\begin{figure}
    \centering
    \includegraphics[width=\columnwidth]{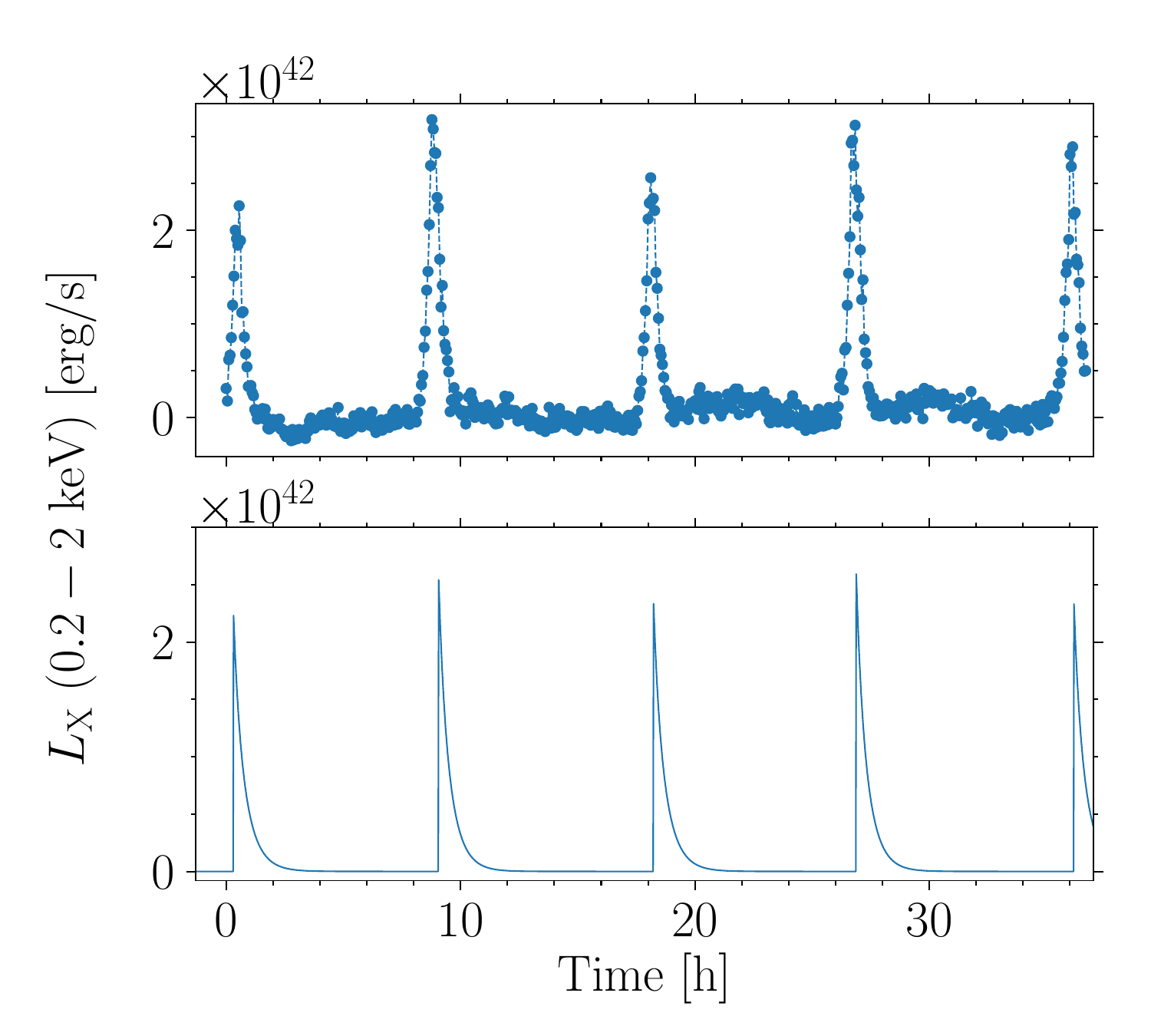}
    \caption{Comparison between observed and predicted light curve. Upper panel: $0.2$-$2$ keV quiescence-subtracted X-ray luminosity light curve from the {\it XMM-Newton} observation XMM5 of GSN 069 \citep{2023A&A...670A..93M}. Lower panel: Synthetic light curve obtained with the parameters given in Sect. \ref{subsection:GSN}.} 
    \label{fig:GSN}
\end{figure}

GSN 069 is the most studied source that has shown QPEs in different epochs. Here, we modelled five peaks observed during an {\it XMM-Newton} observation performed on 31 May 2019, described as XMM5 in \cite{2023A&A...670A..93M}. 
The quiescence bolometric luminosity is $2-5\times10^{43}$ erg s$^{-1}$ for this source, implying $\dot{m}_1\sim 0.1$ for a $10^6M_{\odot}$ primary black hole. We assumed the MBH to be slowly spinning with spin $\chi=0.1$. The secondary BH orbits the primary on a slightly eccentric and inclined orbit ($i_{\rm EMRI}=10^{\circ}$ with respect to the $x-y$ plane), with a semi-major axis of $a=160R_{\rm g}$, giving an orbital period of $18$ hr, eccentricity $e=0.1$, yielding a GW timescale of $T_{\rm gw} \sim 2\times 10^4$ yr. We assumed that the disc contains $4M_{\odot}$, is misaligned by $i_{\rm disc}=5^{\circ}$ with respect to the $x-y$ plane and precesses rigidly with a period of $\sim 125$ days. Based on the detected quiescence level, we assumed $\dot{m}_1 = 0.1$. As shown in Figure~\ref{fig:GSN}, our model nicely reproduces not only the observed alternating long-short recurrence times, but also the associated alternating strong-weak QPEs \citep{2023A&A...670A..93M}. Our model also predicts a peak temperature of $\sim 100$ eV, in agreement with the observed values, namely, 90-95 eV \citep{2023A&A...670A..93M}.

\subsubsection{eRO-QPE2}
\label{subsection:ero2}

For this less massive source, the bolometric luminosity of the quiescence is $\sim 5-6\times 10^{41}$ erg s$^{-1}$, corresponding to $\dot{m}_1\sim 0.01$ for a $M_1=10^5 M_{\odot}$ primary MBH \citep{Arcodia2021}. We assumed the primary to have a spin of $\chi=0.5$.
The secondary BH orbits the primary on an almost circular inclined orbit with a semi-major axis of $a=320R_{\rm g}$, giving an orbital period of $5$ hr,  eccentricity $e=0.05$ (yielding $T_{\rm gw} \sim 3000$ yr) and inclination of $i_{\rm EMRI}=15^{\circ}$ with respect to the $x-y$ plane.

We note that the disc mass in this case is likely too low to be compatible with the tidal disruption of a solar mass star. However, we cannot rule out the possibility that part of the disc mass has already been accreted onto the MBH.

\begin{figure}
    \centering
    \includegraphics[width=\columnwidth]{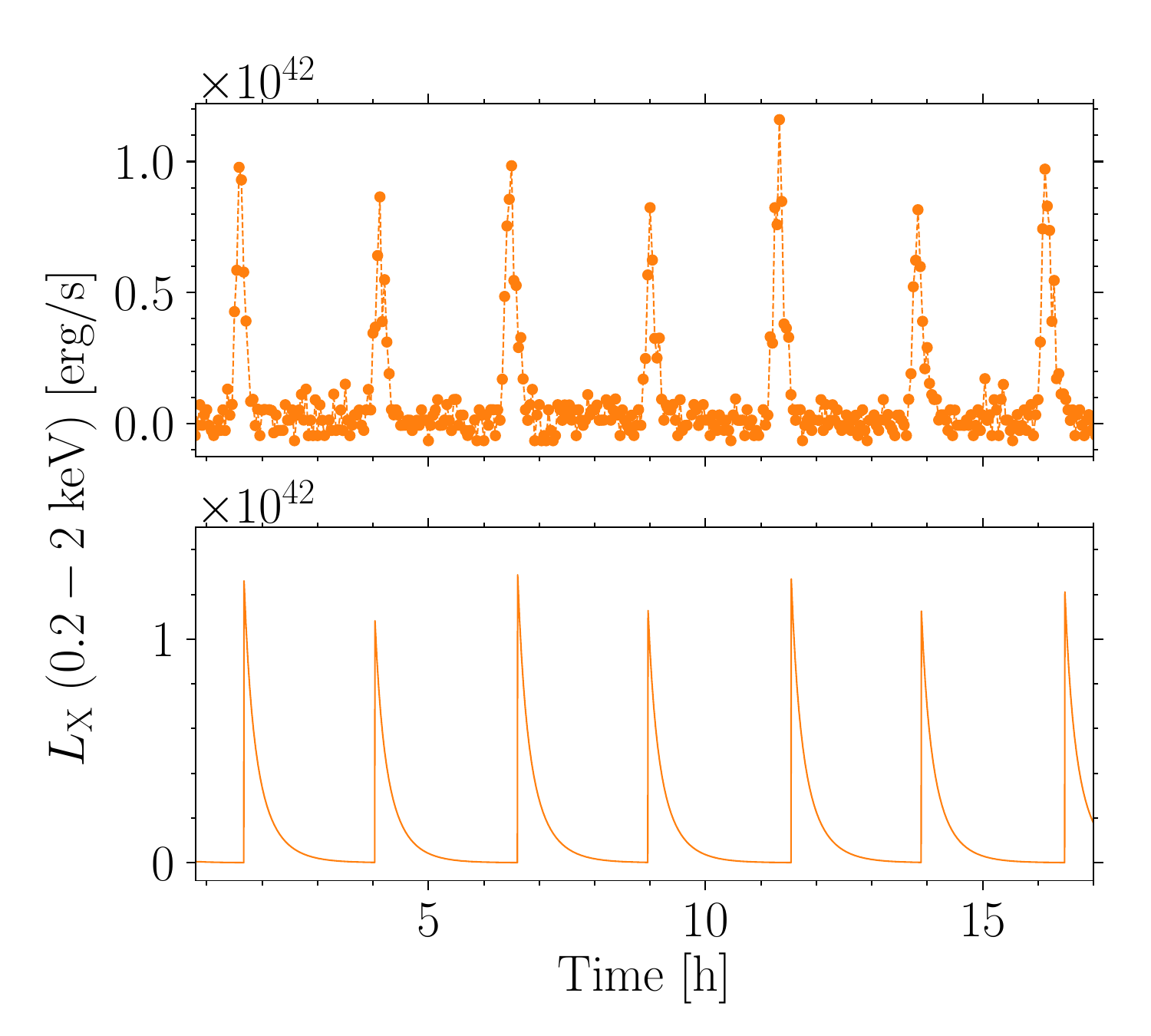}
    \caption{Comparison between observed and predicted light curve. Upper panel: $0.2$-$2$~keV quiescence-subtracted X-ray luminosity light curve from one of the {\it XMM-Newton} observations of eRO-QPE2. Lower panel: Synthetic light curve obtained with the parameters listed in Sect.~\ref{subsection:ero2}.}
    \label{fig:eRO2}
\end{figure}

We assumed that the disc contains $0.01M_{\odot}$ and that is misaligned by $i_{\rm disc}=10^{\circ}$ with respect to the $x-y$ plane, while precessing rigidly on a period of $\sim 5.7$ days. Based on the detected quiescence level, we assumed $\dot{m}_1=0.04$. 
As shown in Figure~\ref{fig:eRO2}, the observed alternating long-short recurrence times and strong-weak QPE amplitudes are remarkably well reproduced, as is the case for GSN~069. The peak temperatures we obtained, namely $\sim 170$ eV, are slightly below the $\sim 215$ eV inferred from observations, although these are, in fact, subject to large uncertainties \citep{Arcodia2021}.

\subsubsection{eRO-QPE1}
\label{sec:ero1}

The quiescence level was not clearly detected in this source, but it is estimated to be below $1.6\times 10^{41}$ erg/s, broadly corresponding to $\dot{m}_1 \sim 0.05-0.001$ for a primary MBH with $M_1=10^{5.8} M_{\odot}$ \citep{Arcodia2021}.
For this source, we assumed the MBH to have a spin of $\chi=0.65$, while the secondary orbit is taken to be slightly eccentric and inclined, with a semi-major axis of $a=355R_{\rm g}$, giving an orbital period of $40$ hr, eccentricity $e=0.05$ (with $T_{\rm gw} \sim 2\times 10^5$ yr) and inclination $i_{\rm EMRI}=20^{\circ}$ with respect to the $x-y$ plane.
We assumed that the disc contains $4M_{\odot}$, is misaligned by $i_{\rm disc}=5^{\circ}$ with respect to the $x-y$ plane and precesses rigidly on a $\sim 7.5$ day timescale. Based on the detected quiescence level, we assumed $\dot{m}_1=0.05$. 

\begin{figure}
    \centering
    \includegraphics[width=\columnwidth]{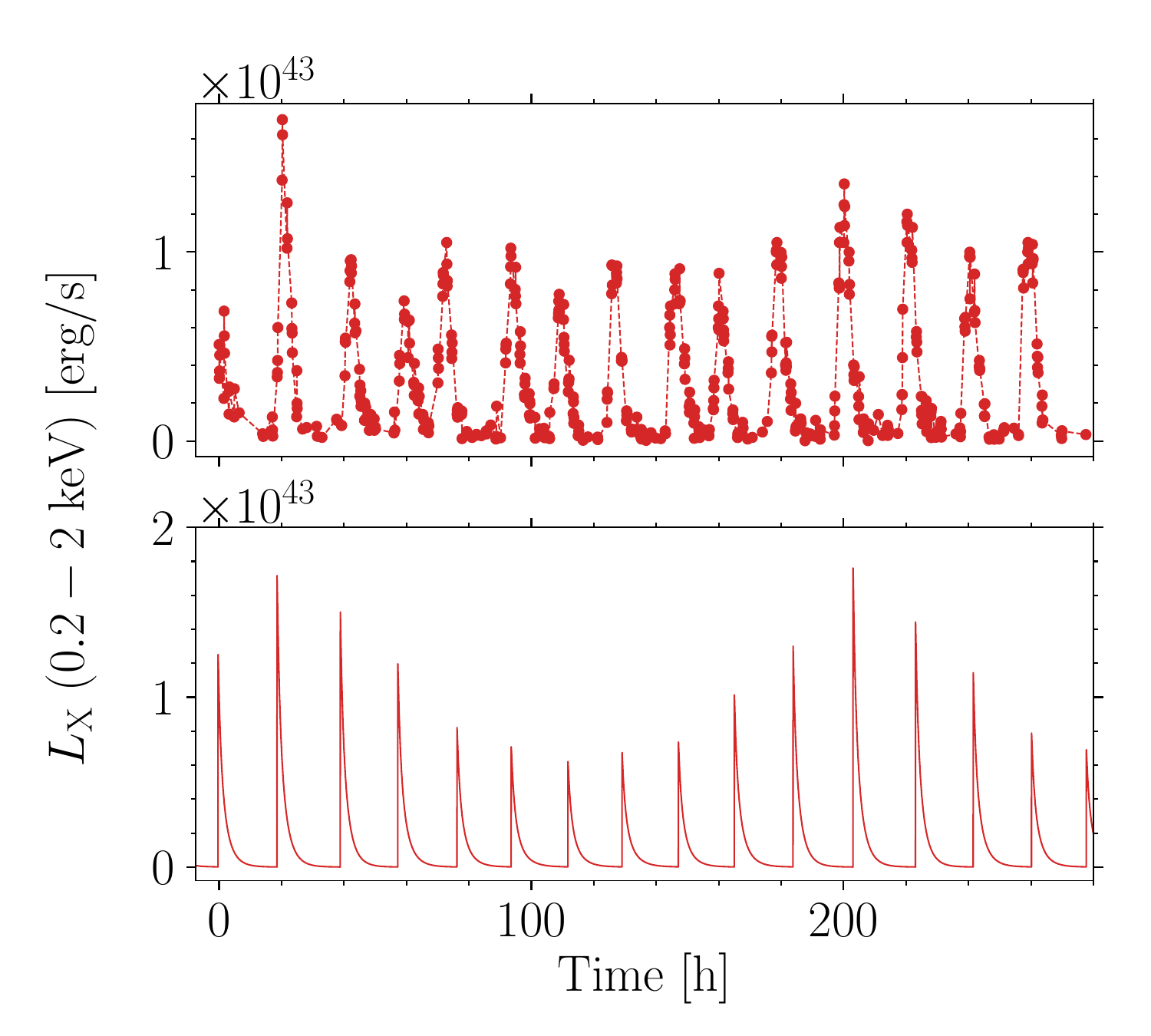}
    \caption{Comparison between observed and predicted light curve. Upper panel: $0.2$-$2$~keV X-ray luminosity light curve from a {\it NICER} $\sim 250$~hr-long monitoring of eRO-QPE1. The quiescent level is undetected by {\it NICER}. Lower panel: Synthetic light curve obtained with the parameters listed in Sect.~\ref{sec:ero1}.}
    \label{fig:eRO1}
\end{figure}
We show the light curve in Figure~\ref{fig:eRO1}.
We note that there is a hint of a superposed modulation on a $\sim 150$ hr period in the data of this QPE source. Having reproduced a similar modulation, we argue that it is linked to the disc precession frequency ($\sim 6$ days).
We inferred a peak temperature of $\sim 190$ eV, broadly consistent with the observed value of $\sim 260$ eV and again subject to large uncertainties \citep{Arcodia2021}.
Here, we note that since our model predicts the QPE occurrence based on the interplay of three different precession frequencies, it could (in principle) reproduce more complex behaviours, such as in the case shown in \cite{Arcodia+2022:ero1_complex}. In particular, those authors found one single QPE peak to be reasonably well fitted by multiple flare profiles (see left panel of their Fig. 2). A more eccentric EMRI could reproduce peaks of different amplitudes much closer in time, according to our model. However, the proper modelisation of the eRO-QPE1 light curve presented in \cite{Arcodia+2022:ero1_complex} warrants a separate, more thorough investigation.

\subsubsection{RX J1301.9+2747}
\label{subsection:RXJ}

RX J1301.9+2747 has a quiescence level similar to that of GSN 069, namely, $\sim 3-6\times 10^{43}$ erg s$^{-1}$, corresponding to $\dot{m}_1\sim 0.1$ for a primary MBH with $M_1=10^{6.3} M_{\odot}$ \citep{Giustini2020}.
We assumed the primary MBH spin to be $\chi=0.5$. The secondary orbits the primary on an eccentric and inclined orbit with a semi-major axis of $a=50R_{\rm g}$, giving an orbital period of $\sim 6$ hr, eccentricity of $e=0.4$ (implying the smallest GW timescale of the four QPEs, i.e. $T_{\rm gw} \sim 400$ yr) and inclination of $i_{\rm EMRI}=6^{\circ}$ with respect to the $x-y$ plane.
We assumed that the disc is characterised by $4M_{\odot}$, misaligned by $i_{\rm disc}=5^{\circ}$ with respect to the $x-y$ plane, and that is precesses rigidly with a period of $\sim 19.8$ days. Based on the detected quiescence level, we assumed $\dot{m}_1=0.5$.
The peak temperature our model predicts is in the range of $\sim 180-450$ eV. This is up to a factor of 3-4 higher than the observed value, namely, 110-130 eV \citep{Giustini2020}.

\begin{figure}
    \centering
    \includegraphics[width=\columnwidth]{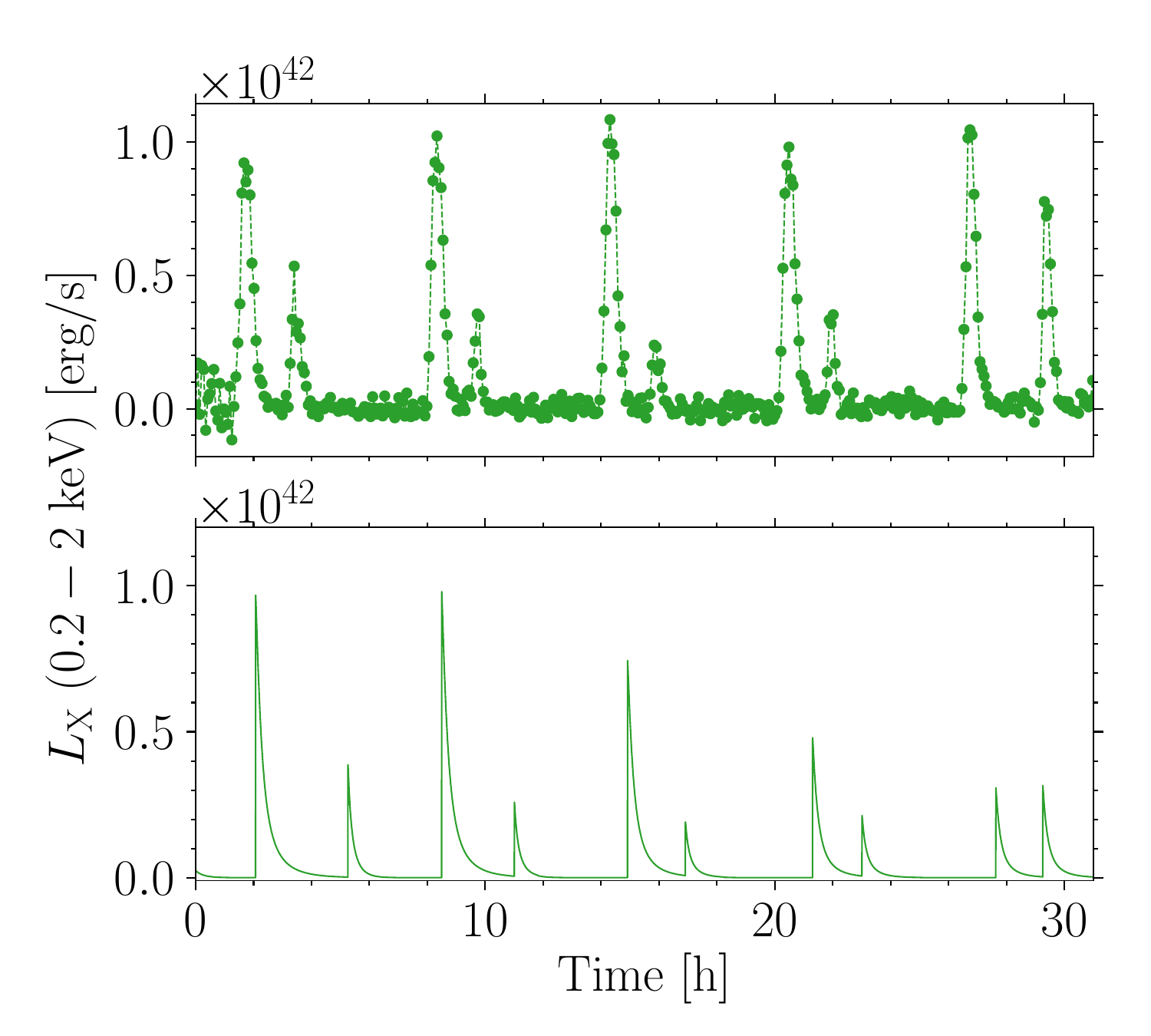}
    \caption{Comparison between observed and predicted light curve. Upper panel: $0.2$-$2$~keV quiescence-subtracted X-ray luminosity from one of the {\it XMM-Newton} observations of RX~J1301.9+2747. Lower panel: Synthetic light curve obtained with the parameters listed in Sect.~\ref{subsection:RXJ}. }
    \label{fig:RXJ}
\end{figure}

Figure~\ref{fig:RXJ} shows that the synthetic QPE amplitudes are compatible with the observed ones within a factor of $\sim 2$ and the complex timing properties of the light curve are reasonably reproduced. However, we note that the observed QPE relative amplitude of the peak pairs is not exactly reproduced by our model. These differences, together with those in the peak temperatures, will be better addressed by both a multi-epoch fitting and a further refinement of the emission mechanism modelisation.

\subsection{Effect of the model free parameters on the light curves}

\begin{figure*}
    \centering
    \includegraphics[width=0.45\textwidth]{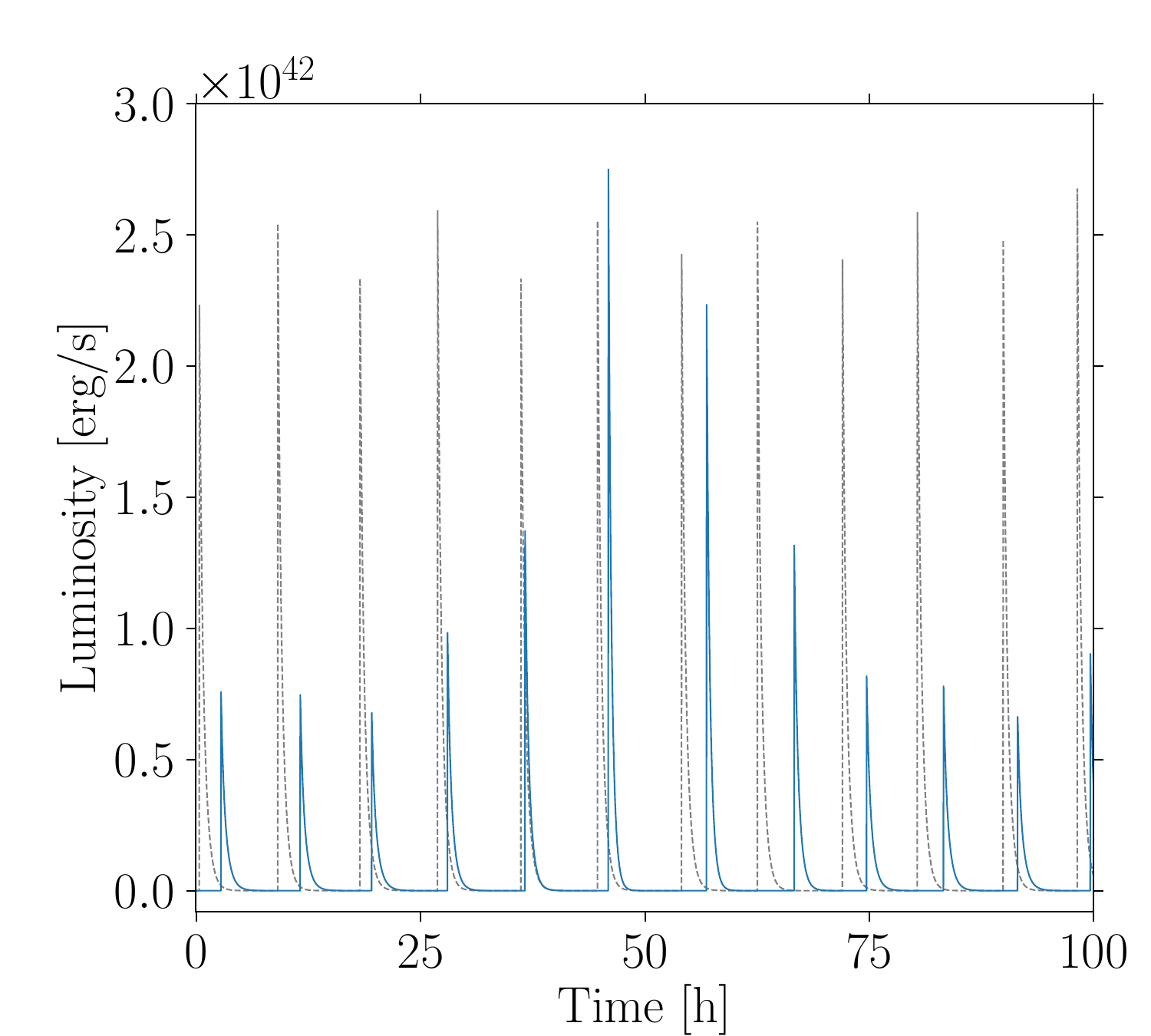}
    \includegraphics[width=0.45\textwidth]{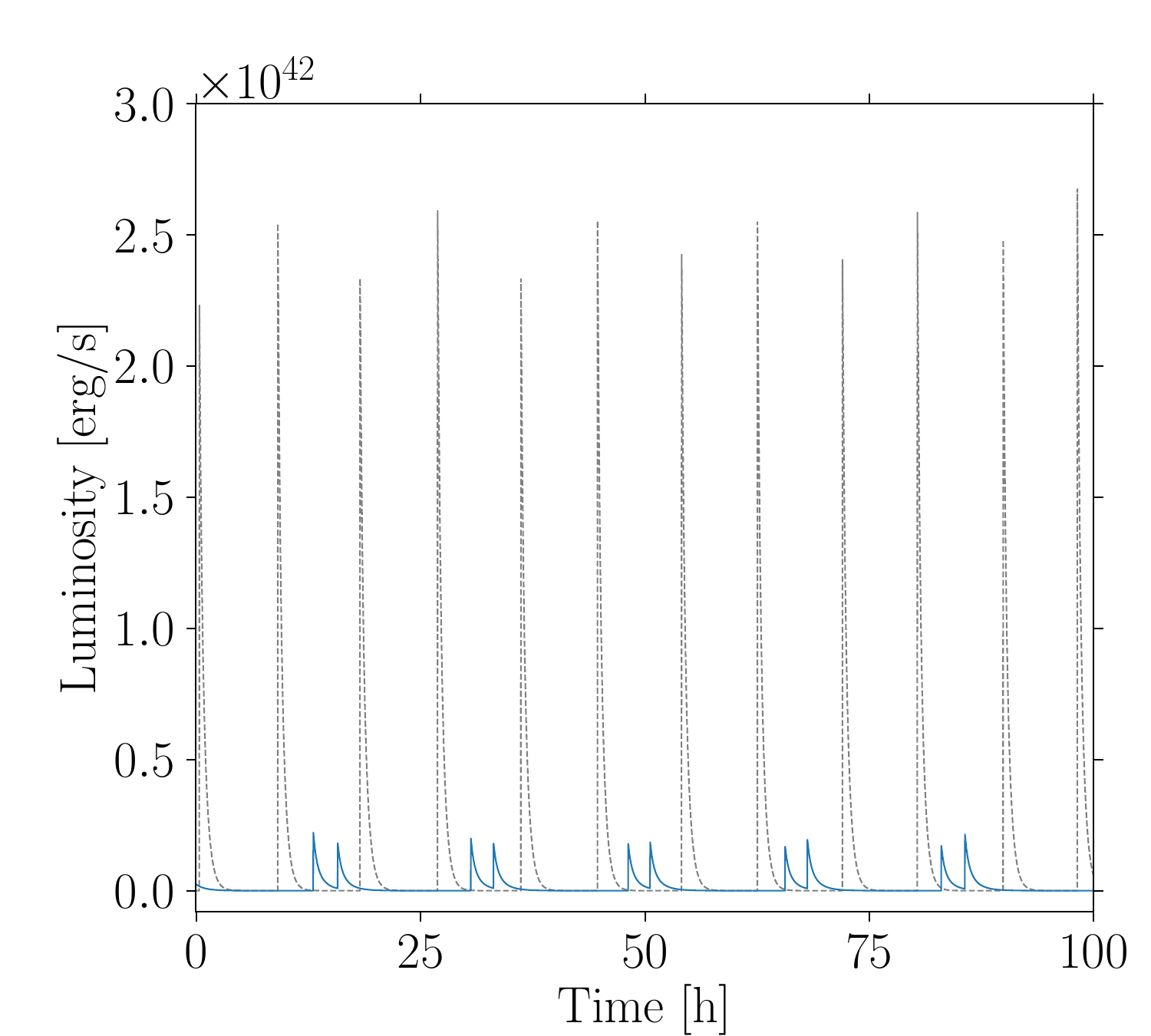}
    \includegraphics[width=0.45\textwidth]{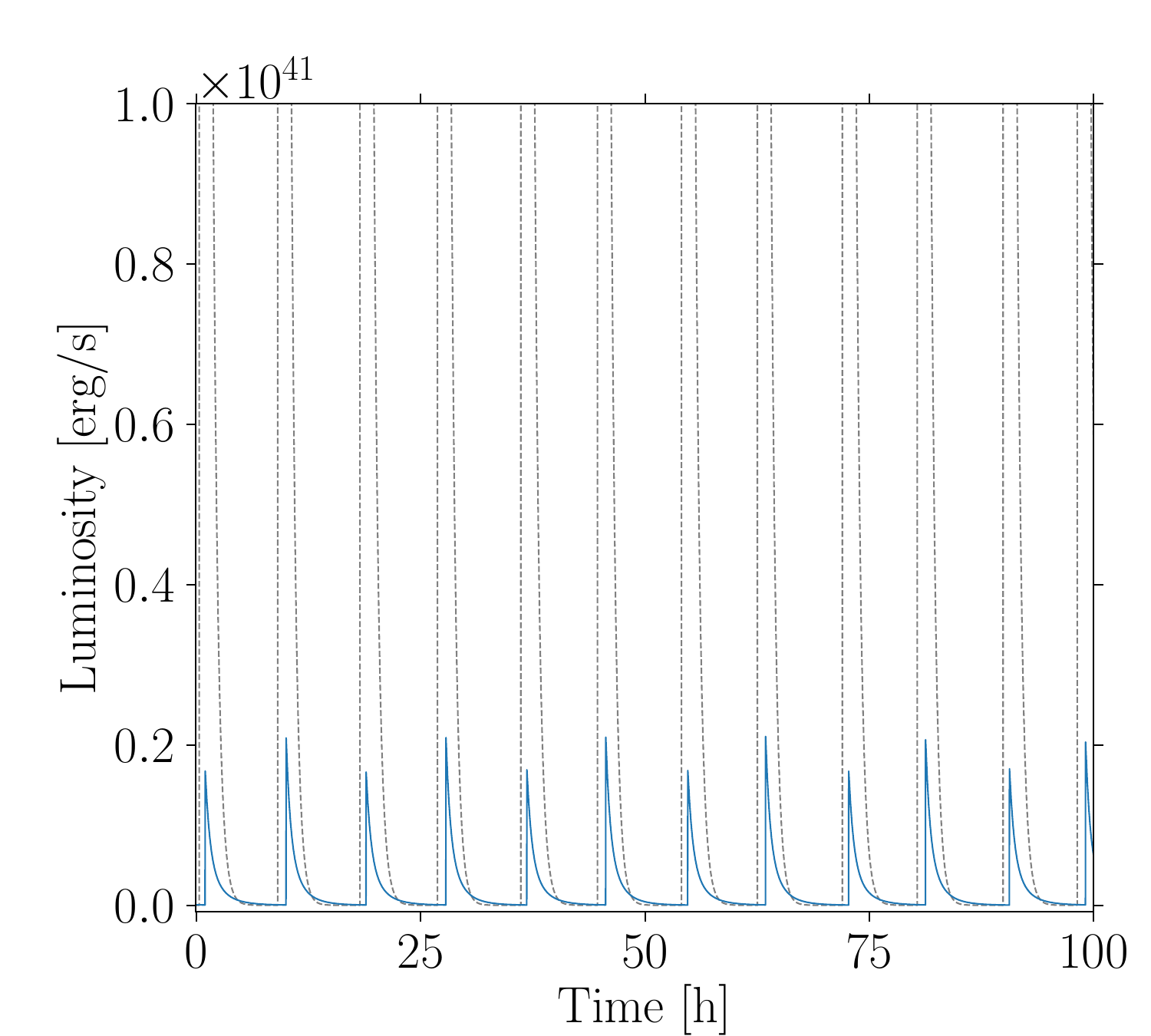}
    \includegraphics[width=0.45\textwidth]{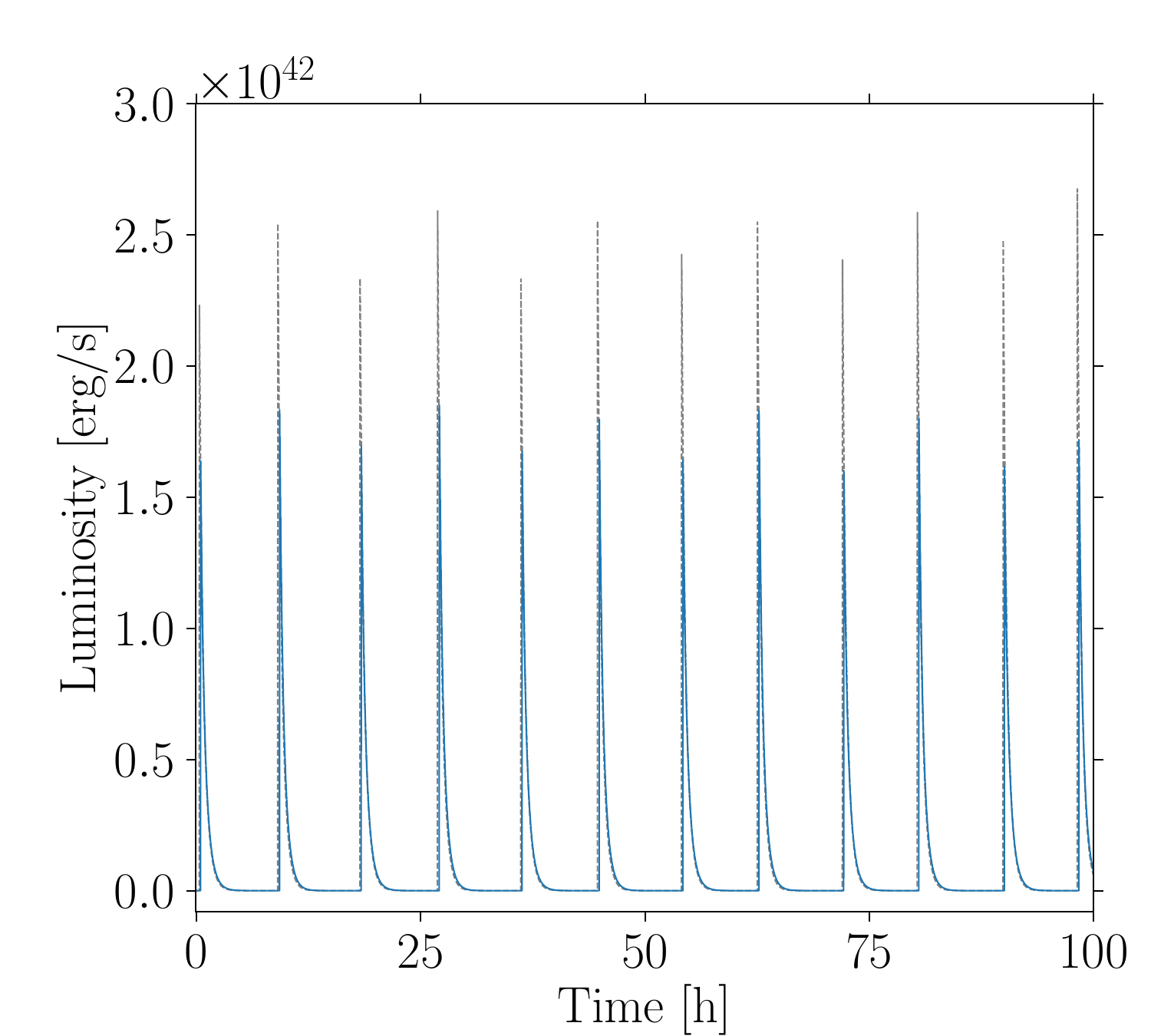}
    \caption{Effect of changing the model parameters on the synthetic light curves for the specific case of GSN 069. We remind that our best-inferred values for the eccentricity of this source is $e=0.1$, MBH spin $\chi=0.1,$ and inclination $i_{\rm EMRI}=10^{\circ}$ (shown by grey dashed lines). In each panel, we only vary one parameter at a time. Top left panel: MBH spin changed to $\chi=0.9$. Top right panel: EMRI eccentricity increased to $e=0.6$. Bottom left panel: Retrograde EMRI with $i_{\rm EMRI}=160^{\circ}$ (note the different $y$-scale). Bottom right panel: Disc precession is switched off.}
    \label{fig:parameter_variation}
\end{figure*}

There are essentially five main parameters that we vary in order to reproduce the synthetic light curves.
Here, we describe the effects of each on the timing and burst properties. We show the main effects in Figure \ref{fig:parameter_variation} for the case of GSN 069. 

The mass of the primary MBH sets the semi-major axis of the EMRI orbit and determines the size of the accretion disc, therefore (together with the MBH spin) the rigid precession period.
The size of the disc is also determined by the primary MBH spin value through the ISCO. In particular the ISCO is smaller, namely, closer to the primary, for higher spin values; therefore, the disc rigidly precesses on shorter timescales. This is because the Lense-Thirring precession frequency at each radius is higher and the average precession frequency increases as the Lense-Thirring precession is stronger at smaller radii. Changing the MBH spin then results in a modulation of the peak amplitude on a timescale that is shorter for higher spins, as shown in the top left panel of Figure \ref{fig:parameter_variation}.
This modulation can also be seen in Figure \ref{fig:eRO1} and corresponds to the disc precession period.

Another parameter that we varied between different sources is the disc mass. This parameter affects the absolute amplitude of the peaks without influencing either the timing or the relative amplitude between subsequent peaks. We  do not explicitly show its effect on the light curve here.

The eccentricity of the EMRI changes substantially the timing properties of the light curve and the amplitude pattern between subsequent peaks. 
Specifically, when the eccentricity is large the crossings near pericentre are closer in time, generating a light curve characterised by closely spaced double peaks separated by larger time intervals. Furthermore, since the eccentric motion causes the binary to change velocity along its orbit, at each crossing, the velocity of the secondary object is quite different from that of the gas in the disc, assumed to rotate in circular motion at all radii. This means that in general eccentric EMRIs are characterised by higher relative velocities, which correspond to lower influence radii and, consequently, lower luminosities.
This is shown in the top right panel of Figure \ref{fig:parameter_variation}.

Changing the relative inclination between the EMRI companion and the disc modifies not only the peak luminosities, but also the relative amplitude of subsequent peaks. 
In particular, if the EMRI companion is orbiting in a retrograde orbit with respect to the disc, the luminosities remain much lower than the observed values as the relative velocity between the EMRI companion and the disc at the crossing is significantly higher (by roughly an order of magnitude) than in the prograde case. This is shown in the bottom left panel of Figure \ref{fig:parameter_variation}.

Finally, we have taken the disc not to be precessing in the case of GSN 069. We find that if the disc position is fixed in time, the peak luminosities are lower by $\sim 30$\%, while the relative amplitude of subsequent peaks decreases by about a factor of 2. We show this in the bottom right panel of Figure \ref{fig:parameter_variation}. The main reason for this change is that the  crossing radii are different compared to the precessing case and this essentially translates into different relative velocities. 

\section{Discussion and conclusions}
\label{sec:concl}

We have built a complete model to reproduce the behaviour of the sources that produce QPEs confirmed to date: GSN 069, eRO-QPE1, eRO-QPE2, and RX J1301.9+2747.
We assumed the QPEs to be triggered when an EMRI companion (mass ratio $q\sim 10^{-4}-10^{-3}$) crosses the tilted and rigidly precessing accretion disc that surrounds the primary MBH. Since QPEs appear to be connected with candidate TDEs (either partial or total), we modelled the accretion disc following \citet{Franchini2016}. 
Therefore, the disc contains a mass on the order of $1 M_{\odot}$ and it is geometrically thick, allowing it to precess as a solid body around the primary.
In our model, the emission comes from a cloud of gas that is pulled out of the disc by the passage of the companion and cools down through adiabatic expansion, emitting $\sim10^{42}-10^{43}$ erg s$^{-1}$ in the soft X-ray band.

We assumed the disc to extend up to $\sim 6\,R_{\rm t}$, where $R_{\rm t}$ is the tidal disruption radius of the primary MBH. This might be an overestimation of the disc's outer radius, as the circularisation radius for the debris of a tidally disrupted star is expected to be $\sim 2\,R_{\rm t}$. However, to date, there have been no complete numerical simulations to validate this assumption and the disc is expected to spread viscously. Viscous spreading of the disc may actually explain why QPEs are only observed with a delay $\gtrsim 4.5$~yr with respect to the initial TDE-like X-ray flare in GSN~069 \citep{2023A&A...670A..93M}, as QPE emission only starts when the outer disc reaches the pre-existing EMRI companion orbit.

We have specifically addressed the case of a BH companion, but we cannot exclude that QPEs are produced by star-disc impacts instead \citep{2023arXiv230316231L,2023arXiv230403670T}, although with the caveats and limitations discussed in Sect.~\ref{sec:secondary}. Once a secondary BH is considered, observable QPEs are only produced in a relatively narrow range of BH masses on the order of $\sim 100$~M$_\odot$ and only for prograde secondary orbits with relatively small inclination between the orbital and disc planes. Significantly lower mass BHs, retrograde secondary orbits, or a large misalignment between the secondary and disc planes all produce X-ray luminosities that are undetectable against the quiescence emission as the influence radius is $R_{\rm inf} \ll 10^{11}$~cm. This is either because of the smaller BH size or because of higher $v_{\rm rel}$ (or both). Moreover, the initial temperature of the cloud would be significantly different and the black body emission peak would not be compatible with the observed $\sim 100$ eV. On the other hand, the EMRI lifetime for significantly higher BH masses could be too short to be reasonably accessible to observations. Therefore, the observed QPE sources might well represent the high-luminosity tail of the QPE population with most systems being actually below the current detection threshold or too short-lived to be detected. In this case, the overall rate of EMRIs could be significantly higher than the one that can be estimated from the observed QPE rate, possibly representing a crucial mechanism for the mass growth of MBHs in the mass range $10^5-10^7$M$_{\odot}$. Furthermore, the fact that even a small number of QPE sources can give a lower limit on the EMRI rates constitutes a relevant prediction for LISA \citep{2023LRR....26....2A}.

We set $\Delta t_{\rm QPE}$ to match the observed duration of the QPEs of each individual source. We do not link the duration to the physical properties of the impact. Therefore our model can (in principle) allow for a wider range of $\Delta t_{\rm QPE}$. The constraints on the duration interval would depend on the physical properties of the expanding cloud of gas as well as on the properties of the gas surrounding the accretion disc. We do, however, note that the duration of the QPE is broadly consistent with the expected expansion velocity that we would get from our theoretical model -- at least for the cases of GSN 069, eRO-QPE1, and eRO-QPE2. We also note that we fixed the expansion factor to $3$ based on the observations of GSN 069 \citep{2023A&A...670A..93M}, as the properties of the gas outside the accretion disc (which determine the timescale for the adiabatic expansion) are unconstrained. However, QPEs become undetectable once the contrast between the cloud temperature and that of the accretion disc emission is $\sim 1$ which implies an expansion factor of $2$-$3$ in the other QPE sources as well. Thus, we refrain from discussing the details of the physical mechanism that drives the expansion of the cloud producing QPEs, deferring a physical treatment to a follow-up paper in which both the QPE rise properties (that we neglect here), decay timescales, and physical properties will be studied in detail. 

We find that only mutual inclinations (i.e. between the EMRI and the disc) smaller than $\sim15^\circ$ and prograde orbits are capable of reproducing the observed luminosities. This would imply, assuming isotropy, that roughly 2\% of the EMRI-disc systems fall into this region of the parameter space.
This is clearly a small fraction of the putative number of possible systems, but considering the scarcity of the sources where QPEs have been identified so far, we believe that this constraint can still be reconciled with current observations.
However, we can reasonably expect higher inclination systems to progressively align the disc outer parts  with the EMRI companion \citep{2013ApJ...774...43M}, therefore transiting through the region of inclinations that would give rise to detectable flares and alleviating a possible tension. 
Finally, since the number of expected EMRI is not very well constrained, we can reasonably expect that for each EMRI detected as QPE there would likely be a population of EMRIs that do not produce QPEs that would be strong enough to be detected.

Another relevant aspect that should be discussed concerns the orientation of the systems considered with respect to the line of sight. Given the primary masses and semi-major axes of the modelled sources, we do not expect timing delays (see Sec.~\ref{sec:crossing}) to be signficant enough to critically change the picture for certain line of sight directions over others. The same can be said for the lensing generated by the MBHs, which (due to the lower masses and relatively wide orbits) seems to produce quite small deflection angles ($\sim 0.01-0.1$ radians in all considered cases) in general. 
Microlensing effects by the orbiting EMRI companions are expected to be even smaller as lensing is only efficient for a very limited range of possible geometrical configurations. We cannot, however, exclude the occasional magnification of one QPE that can occur when the emitting cloud, the MBH, and the observer are close to be perfectly aligned.

The proposed model has five unconstrained parameters: MBH spin value, orbital eccentricity, disc mass, EMRI, and disc inclination. All five may (in principle) be tuned to reproduce the different observed behaviours. 
For simplicity, we have fixed the values of $M_1$ and $M_2$ (and therefore $a$), but all the eight parameters can (in principle) be determined accurately using multi-epoch observations with several QPEs. We aim to build a Bayesian inference algorithm based on our model in order to place meaningful posterior on these parameters from the observed QPE light curve.
This will be the subject of a follow up paper.

Although the emission process we consider is simple and subject to further refinements and modifications, we have shown that our model can reproduce the diversity of QPE recurrence times and behaviour in the four best-studied QPE sources. Moreover, the model is consistent with the observed spectral properties of QPEs and with their evolution during QPE decay and is also able to account naturally for the varying QPE amplitudes and recurrence times in individual erupters.

\begin{acknowledgements}
We thank the anonymous Referee for the useful comments and suggestions.
AF, EB and AS acknowledge financial support provided under the European Union’s H2020 ERC Consolidator Grant ``Binary Massive Black Hole Astrophysics" (B Massive, Grant Agreement: 818691). MB acknowledges support provided by MUR under grant ``PNRR - Missione 4 Istruzione e Ricerca - Componente 2 Dalla Ricerca all'Impresa - Investimento 1.2 Finanziamento di progetti presentati da giovani ricercatori ID:SOE\_0163'' and by University of Milano-Bicocca under grant ``2022-NAZ-0482/B''.
RA received support for this work by NASA through the NASA Einstein Fellowship grant No HF2-51499 awarded by the Space Telescope Science Institute, which is operated by the Association of Universities for Research in Astronomy, Inc., for NASA, under contract NAS5-26555. EB  acknowledges support from  the
European Union’s Horizon 2020 Programme under the AHEAD2020 project (grant
agreement n.~871158). MG is supported by the ``Programa de Atracci\'on de Talento'' of the Comunidad de Madrid, grant number 2018-T1/TIC-11733. GM was partly supported by grant n. PID2020-115325GB-C31 funded by MCIN/AEI/10.13039/501100011033.
\end{acknowledgements}

%
%

\bibliographystyle{aa} 
\bibliography{biblio}

\begin{thebibliography}{50}
\expandafter\ifx\csname natexlab\endcsname\relax\def\natexlab#1{#1}\fi

\bibitem[{{Amaro-Seoane} {et~al.}(2023){Amaro-Seoane}, {Andrews}, {Arca Sedda},
  {Askar}, {Baghi}, {Balasov}, {Bartos}, {Bavera}, {Bellovary}, {Berry},
  {Berti}, {Bianchi}, {Blecha}, {Blondin}, {Bogdanovi{\'c}}, {Boissier},
  {Bonetti}, {Bonoli}, {Bortolas}, {Breivik}, {Capelo}, {Caramete},
  {Cattorini}, {Charisi}, {Chaty}, {Chen}, {Chru{\'s}li{\'n}ska}, {Chua},
  {Church}, {Colpi}, {D'Orazio}, {Danielski}, {Davies}, {Dayal}, {De Rosa},
  {Derdzinski}, {Destounis}, {Dotti}, {Du{\r{A}}{\textsterling}an}, {Dvorkin},
  {Fabj}, {Foglizzo}, {Ford}, {Fouvry}, {Franchini}, {Fragos}, {Fryer},
  {Gaspari}, {Gerosa}, {Graziani}, {Groot}, {Habouzit}, {Haggard}, {Haiman},
  {Han}, {Istrate}, {Johansson}, {Khan}, {Kimpson}, {Kokkotas}, {Kong},
  {Korol}, {Kremer}, {Kupfer}, {Lamberts}, {Larson}, {Lau}, {Liu},
  {Lloyd-Ronning}, {Lodato}, {Lupi}, {Ma}, {Maccarone}, {Mandel}, {Mangiagli},
  {Mapelli}, {Mathis}, {Mayer}, {McGee}, {McKernan}, {Miller}, {Mota},
  {Mumpower}, {Nasim}, {Nelemans}, {Noble}, {Pacucci}, {Panessa},
  {Paschalidis}, {Pfister}, {Porquet}, {Quenby}, {Ricarte}, {R{\"o}pke},
  {Regan}, {Rosswog}, {Ruiter}, {Ruiz}, {Runnoe}, {Schneider}, {Schnittman},
  {Secunda}, {Sesana}, {Seto}, {Shao}, {Shapiro}, {Sopuerta}, {Stone},
  {Suvorov}, {Tamanini}, {Tamfal}, {Tauris}, {Temmink}, {Tomsick}, {Toonen},
  {Torres-Orjuela}, {Toscani}, {Tsokaros}, {Unal}, {V{\'a}zquez-Aceves},
  {Valiante}, {van Putten}, {van Roestel}, {Vignali}, {Volonteri}, {Wu},
  {Younsi}, {Yu}, {Zane}, {Zwick}, {Antonini}, {Baibhav}, {Barausse}, {Bonilla
  Rivera}, {Branchesi}, {Branduardi-Raymont}, {Burdge}, {Chakraborty},
  {Cuadra}, {Dage}, {Davis}, {de Mink}, {Decarli}, {Doneva}, {Escoffier},
  {Gandhi}, {Haardt}, {Lousto}, {Nissanke}, {Nordhaus}, {O'Shaughnessy},
  {Portegies Zwart}, {Pound}, {Schussler}, {Sergijenko}, {Spallicci},
  {Vernieri}, \& {Vigna-G{\'o}mez}}]{2023LRR....26....2A}
{Amaro-Seoane}, P., {Andrews}, J., {Arca Sedda}, M., {et~al.} 2023, Living
  Reviews in Relativity, 26, 2

\bibitem[{{Arcodia} {et~al.}(2021){Arcodia}, {Merloni}, {Nandra}, {Buchner},
  {Salvato}, {Pasham}, {Remillard}, {Comparat}, {Lamer}, {Ponti}, {Malyali},
  {Wolf}, {Arzoumanian}, {Bogensberger}, {Buckley}, {Gendreau}, {Gromadzki},
  {Kara}, {Krumpe}, {Markwardt}, {Ramos-Ceja}, {Rau}, {Schramm}, \&
  {Schwope}}]{Arcodia2021}
{Arcodia}, R., {Merloni}, A., {Nandra}, K., {et~al.} 2021, \nat, 592, 704

\bibitem[{{Arcodia} {et~al.}(2022){Arcodia}, {Miniutti}, {Ponti}, {Buchner},
  {Giustini}, {Merloni}, {Nandra}, {Vincentelli}, {Kara}, {Salvato}, \&
  {Pasham}}]{Arcodia+2022:ero1_complex}
{Arcodia}, R., {Miniutti}, G., {Ponti}, G., {et~al.} 2022, \aap, 662, A49

\bibitem[{{Arnett}(1980)}]{1980ApJ...237..541A}
{Arnett}, W.~D. 1980, \apj, 237, 541

\bibitem[{{Bahcall} \& {Wolf}(1977)}]{1977ApJ...216..883B}
{Bahcall}, J.~N. \& {Wolf}, R.~A. 1977, \apj, 216, 883

\bibitem[{{Bardeen} {et~al.}(1972){Bardeen}, {Press}, \&
  {Teukolsky}}]{1972ApJ...178..347B}
{Bardeen}, J.~M., {Press}, W.~H., \& {Teukolsky}, S.~A. 1972, \apj, 178, 347

\bibitem[{{Blanchet}(2014)}]{Blanchet2014}
{Blanchet}, L. 2014, Living Reviews in Relativity, 17, 2

\bibitem[{{Boh{\'e}} {et~al.}(2013){Boh{\'e}}, {Marsat}, {Faye}, \&
  {Blanchet}}]{2013CQGra..30g5017B}
{Boh{\'e}}, A., {Marsat}, S., {Faye}, G., \& {Blanchet}, L. 2013, Classical and
  Quantum Gravity, 30, 075017

\bibitem[{{Bonetti} {et~al.}(2016){Bonetti}, {Haardt}, {Sesana}, \&
  {Barausse}}]{2016MNRAS.461.4419B}
{Bonetti}, M., {Haardt}, F., {Sesana}, A., \& {Barausse}, E. 2016, \mnras, 461,
  4419

\bibitem[{{Bortolas}(2022)}]{2022MNRAS.511.2885B}
{Bortolas}, E. 2022, \mnras, 511, 2885

\bibitem[{{Broggi} {et~al.}(2022){Broggi}, {Bortolas}, {Bonetti}, {Sesana}, \&
  {Dotti}}]{2022MNRAS.514.3270B}
{Broggi}, L., {Bortolas}, E., {Bonetti}, M., {Sesana}, A., \& {Dotti}, M. 2022,
  \mnras, 514, 3270

\bibitem[{{Chakraborty} {et~al.}(2021){Chakraborty}, {Kara}, {Masterson},
  {Giustini}, {Miniutti}, \& {Saxton}}]{Chakraborty2021}
{Chakraborty}, J., {Kara}, E., {Masterson}, M., {et~al.} 2021, \apjl, 921, L40

\bibitem[{{Chen} {et~al.}(2022){Chen}, {Qiu}, {Li}, \&
  {Liu}}]{2022ApJ...930..122C}
{Chen}, X., {Qiu}, Y., {Li}, S., \& {Liu}, F.~K. 2022, \apj, 930, 122

\bibitem[{{Franchini} {et~al.}(2016){Franchini}, {Lodato}, \&
  {Facchini}}]{Franchini2016}
{Franchini}, A., {Lodato}, G., \& {Facchini}, S. 2016, \mnras, 455, 1946

\bibitem[{{French} {et~al.}(2020){French}, {Wevers}, {Law-Smith}, {Graur}, \&
  {Zabludoff}}]{2020SSRv..216...32F}
{French}, K.~D., {Wevers}, T., {Law-Smith}, J., {Graur}, O., \& {Zabludoff},
  A.~I. 2020, \ssr, 216, 32

\bibitem[{{Giustini} {et~al.}(2020){Giustini}, {Miniutti}, \&
  {Saxton}}]{Giustini2020}
{Giustini}, M., {Miniutti}, G., \& {Saxton}, R.~D. 2020, \aap, 636, L2

\bibitem[{{Hammerstein} {et~al.}(2021){Hammerstein}, {Gezari}, {van Velzen},
  {Cenko}, {Roth}, {Ward}, {Frederick}, {Hung}, {Graham}, {Foley}, {Bellm},
  {Cannella}, {Drake}, {Kupfer}, {Laher}, {Mahabal}, {Masci}, {Riddle},
  {Rojas-Bravo}, \& {Smith}}]{2021ApJ...908L..20H}
{Hammerstein}, E., {Gezari}, S., {van Velzen}, S., {et~al.} 2021, \apjl, 908,
  L20

\bibitem[{{Ingram} {et~al.}(2021){Ingram}, {Motta}, {Aigrain}, \&
  {Karastergiou}}]{2021MNRAS.503.1703I}
{Ingram}, A., {Motta}, S.~E., {Aigrain}, S., \& {Karastergiou}, A. 2021,
  \mnras, 503, 1703

\bibitem[{{Kaur} {et~al.}(2022){Kaur}, {Stone}, \& {Gilbaum}}]{Kaur2022}
{Kaur}, K., {Stone}, N.~C., \& {Gilbaum}, S. 2022, arXiv e-prints,
  arXiv:2211.00704

\bibitem[{{King}(2020)}]{2020MNRAS.493L.120K}
{King}, A. 2020, \mnras, 493, L120

\bibitem[{{King}(2022)}]{2022MNRAS.515.4344K}
{King}, A. 2022, \mnras, 515, 4344

\bibitem[{{Krolik} \& {Linial}(2022)}]{2022ApJ...941...24K}
{Krolik}, J.~H. \& {Linial}, I. 2022, \apj, 941, 24

\bibitem[{{Kroupa}(2001)}]{2001MNRAS.322..231K}
{Kroupa}, P. 2001, \mnras, 322, 231

\bibitem[{{Lense} \& {Thirring}(1918)}]{Lense1918}
{Lense}, J. \& {Thirring}, H. 1918, Physikalische Zeitschrift, 19, 156

\bibitem[{{Linial} \& {Metzger}(2023)}]{2023arXiv230316231L}
{Linial}, I. \& {Metzger}, B.~D. 2023, arXiv e-prints, arXiv:2303.16231

\bibitem[{{Linial} \& {Sari}(2023)}]{2023ApJ...945...86L}
{Linial}, I. \& {Sari}, R. 2023, \apj, 945, 86

\bibitem[{{Lu} \& {Quataert}(2022)}]{2022arXiv221008023L}
{Lu}, W. \& {Quataert}, E. 2022, arXiv e-prints, arXiv:2210.08023

\bibitem[{{Martin} \& {Franchini}(2021)}]{Martin2021}
{Martin}, R.~G. \& {Franchini}, A. 2021, \apjl, 922, L37

\bibitem[{{Metzger}(2022)}]{2022ApJ...937L..12M}
{Metzger}, B.~D. 2022, \apjl, 937, L12

\bibitem[{{Miller} \& {Krolik}(2013)}]{2013ApJ...774...43M}
{Miller}, M.~C. \& {Krolik}, J.~H. 2013, \apj, 774, 43

\bibitem[{{Miniutti} {et~al.}(2023{\natexlab{a}}){Miniutti}, {Giustini},
  {Arcodia}, {Saxton}, {Chakraborty}, {Read}, \& {Kara}}]{2023A&A...674L...1M}
{Miniutti}, G., {Giustini}, M., {Arcodia}, R., {et~al.} 2023{\natexlab{a}},
  \aap, 674, L1

\bibitem[{{Miniutti} {et~al.}(2023{\natexlab{b}}){Miniutti}, {Giustini},
  {Arcodia}, {Saxton}, {Read}, {Bianchi}, \& {Alexander}}]{2023A&A...670A..93M}
{Miniutti}, G., {Giustini}, M., {Arcodia}, R., {et~al.} 2023{\natexlab{b}},
  \aap, 670, A93

\bibitem[{{Miniutti} {et~al.}(2019){Miniutti}, {Saxton}, {Giustini},
  {Alexander}, {Fender}, {Heywood}, {Monageng}, {Coriat}, {Tzioumis}, {Read},
  {Knigge}, {Gandhi}, {Pretorius}, \& {Ag{\'\i}s-Gonz{\'a}lez}}]{Miniutti2019}
{Miniutti}, G., {Saxton}, R.~D., {Giustini}, M., {et~al.} 2019, \nat, 573, 381

\bibitem[{{Nayakshin} {et~al.}(2004){Nayakshin}, {Cuadra}, \&
  {Sunyaev}}]{2004A&A...413..173N}
{Nayakshin}, S., {Cuadra}, J., \& {Sunyaev}, R. 2004, \aap, 413, 173

\bibitem[{{Pan} {et~al.}(2022){Pan}, {Li}, {Cao}, {Miniutti}, \&
  {Gu}}]{Pan2022}
{Pan}, X., {Li}, S.-L., {Cao}, X., {Miniutti}, G., \& {Gu}, M. 2022, \apjl,
  928, L18

\bibitem[{{Pihajoki}(2016)}]{2016MNRAS.457.1145P}
{Pihajoki}, P. 2016, \mnras, 457, 1145

\bibitem[{{Poisson} \& {Will}(2014)}]{2014grav.book.....P}
{Poisson}, E. \& {Will}, C.~M. 2014, {Gravity}

\bibitem[{{Quintin} {et~al.}(2023){Quintin}, {Webb}, {Guillot}, {Miniutti},
  {Kammoun}, {Giustini}, {Arcodia}, {Soucail}, {Clerc}, {Amato}, \&
  {Markwardt}}]{Quintin2023}
{Quintin}, E., {Webb}, N.~A., {Guillot}, S., {et~al.} 2023, arXiv e-prints,
  arXiv:2306.00438

\bibitem[{{Raj} \& {Nixon}(2021)}]{Raj2021}
{Raj}, A. \& {Nixon}, C.~J. 2021, \apj, 909, 82

\bibitem[{{Shakura} \& {Sunyaev}(1973)}]{SS1973}
{Shakura}, N.~I. \& {Sunyaev}, R.~A. 1973, \aap, 24, 337

\bibitem[{{Stone} \& {Loeb}(2012)}]{SL2012}
{Stone}, N. \& {Loeb}, A. 2012, \prl, 108, 061302

\bibitem[{{Strubbe} \& {Quataert}(2009)}]{Strubbe2009}
{Strubbe}, L.~E. \& {Quataert}, E. 2009, \mnras, 400, 2070

\bibitem[{{Sukov{\'a}} {et~al.}(2021){Sukov{\'a}}, {Zaja{\v{c}}ek}, {Witzany},
  \& {Karas}}]{2021ApJ...917...43S}
{Sukov{\'a}}, P., {Zaja{\v{c}}ek}, M., {Witzany}, V., \& {Karas}, V. 2021,
  \apj, 917, 43

\bibitem[{{Tagawa} \& {Haiman}(2023)}]{2023arXiv230403670T}
{Tagawa}, H. \& {Haiman}, Z. 2023, arXiv e-prints, arXiv:2304.03670

\bibitem[{{The LIGO Scientific Collaboration} {et~al.}(2021){The LIGO
  Scientific Collaboration}, {the Virgo Collaboration}, {the KAGRA
  Collaboration}, {Abbott}, {Abbott}, {Acernese}, {Ackley}, {Adams},
  {Adhikari}, {Adhikari}, {Adya}, {Affeldt}, {Agarwal}, {Agathos}, {Agatsuma},
  {Aggarwal}, {Aguiar}, {Aiello}, {Ain}, {Ajith}, {Akcay}, {Akutsu},
  {Albanesi}, {Allocca}, {Altin}, {Amato}, {Anand}, {Anand}, {Ananyeva},
  {Anderson}, {Anderson}, {Ando}, {Andrade}, {Andres}, {Andri{\'c}},
  {Angelova}, {Ansoldi}, {Antelis}, {Antier}, {Appert}, {Arai}, {Arai}, {Arai},
  {Araki}, {Araya}, {Araya}, {Areeda}, {Ar{\`e}ne}, {Aritomi}, {Arnaud},
  {Arogeti}, {Aronson}, {Arun}, {Asada}, {Asali}, {Ashton}, {Aso}, {Assiduo},
  {Aston}, {Astone}, {Aubin}, {Austin}, {Babak}, {Badaracco}, {Bader},
  {Badger}, {Bae}, {Bae}, {Baer}, {Bagnasco}, {Bai}, {Baiotti}, {Baird},
  {Bajpai}, {Ball}, {Ballardin}, {Ballmer}, {Balsamo}, {Baltus}, {Banagiri},
  {Bankar}, {Barayoga}, {Barbieri}, {Barish}, {Barker}, {Barneo}, {Barone},
  {Barr}, {Barsotti}, {Barsuglia}, {Barta}, {Bartlett}, {Barton}, {Bartos},
  {Bassiri}, {Basti}, {Bawaj}, {Bayley}, {Baylor}, {Bazzan}, {B{\'e}csy},
  {Bedakihale}, {Bejger}, {Belahcene}, {Benedetto}, {Beniwal}, {Bennett},
  {Bentley}, {BenYaala}, {Bergamin}, {Berger}, {Bernuzzi}, {Berry},
  {Bersanetti}, {Bertolini}, {Betzwieser}, {Beveridge}, {Bhandare}, {Bhardwaj},
  {Bhattacharjee}, {Bhaumik}, {Bilenko}, {Billingsley}, {Bini}, {Birney},
  {Birnholtz}, {Biscans}, {Bischi}, {Biscoveanu}, {Bisht}, {Biswas}, {Bitossi},
  {Bizouard}, {Blackburn}, {Blair}, {Blair}, {Blair}, {Bobba}, {Bode}, {Boer},
  {Bogaert}, {Boldrini}, {Bonavena}, {Bondu}, {Bonilla}, {Bonnand}, {Booker},
  {Boom}, {Bork}, {Boschi}, {Bose}, {Bose}, {Bossilkov}, {Boudart},
  {Bouffanais}, {Bozzi}, {Bradaschia}, {Brady}, {Bramley}, {Branch},
  {Branchesi}, {Brandt}, {Brau}, {Breschi}, {Briant}, {Briggs}, {Brillet},
  {Brinkmann}, {Brockill}, {Brooks}, {Brooks}, {Brown}, {Brunett}, {Bruno},
  {Bruntz}, {Bryant}, {Bulik}, {Bulten}, {Buonanno}, {Buscicchio}, {Buskulic},
  {Buy}, {Byer}, {Cabourn Davies}, {Cadonati}, {Cagnoli}, {Cahillane},
  {Calder{\'o}n Bustillo}, {Callaghan}, {Callister}, {Calloni}, {Cameron},
  {Camp}, {Canepa}, {Canevarolo}, {Cannavacciuolo}, {Cannon}, {Cao}, {Cao},
  {Capocasa}, {Capote}, {Carapella}, {Carbognani}, {Carlin}, {Carney},
  {Carpinelli}, {Carrillo}, {Carullo}, {Carver}, {Casanueva Diaz}, {Casentini},
  {Castaldi}, {Caudill}, {Cavagli{\`a}}, {Cavalier}, {Cavalieri}, {Ceasar},
  {Cella}, {Cerd{\'a}-Dur{\'a}n}, {Cesarini}, {Chaibi}, {Chakravarti},
  {Chalathadka Subrahmanya}, {Champion}, {Chan}, {Chan}, {Chan}, {Chan},
  {Chan}, {Chandra}, {Chanial}, {Chao}, {Chapman-Bird}, {Charlton}, {Chase},
  {Chassande-Mottin}, {Chatterjee}, {Chatterjee}, {Chatterjee}, {Chaturvedi},
  {Chaty}, {Chatziioannou}, {Chen}, {Chen}, {Chen}, {Chen}, {Chen}, {Chen},
  {Chen}, {Chen}, {Cheng}, {Cheong}, {Cheung}, {Chia}, {Chiadini}, {Chiang},
  {Chiarini}, {Chierici}, {Chincarini}, {Chiofalo}, {Chiummo}, {Cho}, {Cho},
  {Choudhary}, {Choudhary}, {Christensen}, {Chu}, {Chu}, {Chu}, {Chua},
  {Chung}, {Ciani}, {Ciecielag}, {Cie{\'s}lar}, {Cifaldi}, {Ciobanu}, {Ciolfi},
  {Cipriano}, {Cirone}, {Clara}, {Clark}, {Clark}, {Clarke}, {Clearwater},
  {Clesse}, {Cleva}, {Coccia}, {Codazzo}, {Cohadon}, {Cohen}, {Cohen},
  {Colleoni}, {Collette}, {Colombo}, {Colpi}, {Compton}, {Constancio}, {Conti},
  {Cooper}, {Corban}, {Corbitt}, {Cordero-Carri{\'o}n}, {Corezzi}, {Corley},
  {Cornish}, {Corre}, {Corsi}, {Cortese}, {Costa}, {Cotesta}, {Coughlin},
  {Coulon}, {Countryman}, {Cousins}, {Couvares}, {Coward}, {Cowart}, {Coyne},
  {Coyne}, {Creighton}, {Creighton}, {Criswell}, {Croquette}, {Crowder},
  {Cudell}, {Cullen}, {Cumming}, {Cummings}, {Cunningham}, {Cuoco},
  {Cury{\l}o}, {Dabadie}, {Dal Canton}, {Dall'Osso}, {D{\'a}lya}, {Dana},
  {DaneshgaranBajastani}, {D'Angelo}, {Danila}, {Danilishin}, {D'Antonio},
  {Danzmann}, {Darsow-Fromm}, {Dasgupta}, {Datrier}, {Datta}, {Dattilo},
  {Dave}, {Davier}, {Davis}, {Davis}, {Daw}, {de Alarc{\'o}n}, {Dean}, {DeBra},
  {Deenadayalan}, {Degallaix}, {De Laurentis}, {Del{\'e}glise}, {Del Favero},
  {De Lillo}, {De Lillo}, {Del Pozzo}, {DeMarchi}, {De Matteis}, {D'Emilio},
  {Demos}, {Dent}, {Depasse}, {De Pietri}, {De Rosa}, {De Rossi}, {DeSalvo},
  {De Simone}, {Dhurandhar}, {D{\'\i}az}, {Diaz-Ortiz}, {Didio}, {Dietrich},
  {Di Fiore}, {Di Fronzo}, {Di Giorgio}, {Di Giovanni}, {Di Giovanni}, {Di
  Girolamo}, {Di Lieto}, {Ding}, {Di Pace}, {Di Palma}, {Di Renzo},
  {Divakarla}, {Dmitriev}, {Doctor}, {D'Onofrio}, {Donovan}, {Dooley},
  {Doravari}, {Dorrington}, {Drago}, {Driggers}, {Drori}, {Ducoin}, {Dupej},
  {Durante}, {D'Urso}, {Duverne}, {Dwyer}, {Eassa}, {Easter}, {Ebersold},
  {Eckhardt}, {Eddolls}, {Edelman}, {Edo}, {Edy}, {Effler}, {Eguchi},
  {Eichholz}, {Eikenberry}, {Eisenmann}, {Eisenstein}, {Ejlli}, {Engelby},
  {Enomoto}, {Errico}, {Essick}, {Estell{\'e}s}, {Estevez}, {Etienne}, {Etzel},
  {Evans}, {Evans}, {Ewing}, {Fafone}, {Fair}, {Fairhurst}, {Farah}, {Farinon},
  {Farr}, {Farr}, {Farrow}, {Fauchon-Jones}, {Favaro}, {Favata}, {Fays},
  {Fazio}, {Feicht}, {Fejer}, {Fenyvesi}, {Ferguson}, {Fernandez-Galiana},
  {Ferrante}, {Ferreira}, {Fidecaro}, {Figura}, {Fiori}, {Fishbach}, {Fisher},
  {Fittipaldi}, {Fiumara}, {Flaminio}, {Floden}, {Fong}, {Font}, {Fornal},
  {Forsyth}, {Franke}, {Frasca}, {Frasconi}, {Frederick}, {Freed}, {Frei},
  {Freise}, {Frey}, {Fritschel}, {Frolov}, {Fronz{\'e}}, {Fujii}, {Fujikawa},
  {Fukunaga}, {Fukushima}, {Fulda}, {Fyffe}, {Gabbard}, {Gabella}, {Gadre},
  {Gair}, {Gais}, {Galaudage}, {Gamba}, {Ganapathy}, {Ganguly}, {Gao},
  {Gaonkar}, {Garaventa}, {Garc{\'\i}a}, {Garc{\'\i}a-N{\'u}{\~n}ez},
  {Garc{\'\i}a-Quir{\'o}s}, {Garufi}, {Gateley}, {Gaudio}, {Gayathri}, {Ge},
  {Gemme}, {Gennai}, {George}, {George}, {Gerberding}, {Gergely}, {Gewecke},
  {Ghonge}, {Ghosh}, {Ghosh}, {Ghosh}, {Ghosh}, {Giacomazzo}, {Giacoppo},
  {Giaime}, {Giardina}, {Gibson}, {Gier}, {Giesler}, {Giri}, {Gissi},
  {Glanzer}, {Gleckl}, {Godwin}, {Goetz}, {Goetz}, {Gohlke}, {Golomb},
  {Goncharov}, {Gonz{\'a}lez}, {Gopakumar}, {Gosselin}, {Gouaty}, {Gould},
  {Grace}, {Grado}, {Granata}, {Granata}, {Grant}, {Gras}, {Grassia}, {Gray},
  {Gray}, {Greco}, {Green}, {Green}, {Gretarsson}, {Gretarsson}, {Griffith},
  {Griffiths}, {Griggs}, {Grignani}, {Grimaldi}, {Grimm}, {Grote}, {Grunewald},
  {Gruning}, {Guerra}, {Guidi}, {Guimaraes}, {Guix{\'e}}, {Gulati}, {Guo},
  {Guo}, {Gupta}, {Gupta}, {Gupta}, {Gustafson}, {Gustafson}, {Guzman}, {Ha},
  {Haegel}, {Hagiwara}, {Haino}, {Halim}, {Hall}, {Hamilton}, {Hammond}, {Han},
  {Haney}, {Hanks}, {Hanna}, {Hannam}, {Hannuksela}, {Hansen}, {Hansen},
  {Hanson}, {Harder}, {Hardwick}, {Haris}, {Harms}, {Harry}, {Harry},
  {Hartwig}, {Hasegawa}, {Haskell}, {Hasskew}, {Haster}, {Hattori}, {Haughian},
  {Hayakawa}, {Hayama}, {Hayes}, {Healy}, {Heidmann}, {Heidt}, {Heintze},
  {Heinze}, {Heinzel}, {Heitmann}, {Hellman}, {Hello}, {Helmling-Cornell},
  {Hemming}, {Hendry}, {Heng}, {Hennes}, {Hennig}, {Hennig}, {Hernandez},
  {Hernandez Vivanco}, {Heurs}, {Hild}, {Hill}, {Himemoto}, {Hines},
  {Hiranuma}, {Hirata}, {Hirose}, {Hochheim}, {Hofman}, {Hohmann}, {Holcomb},
  {Holland}, {Holley-Bockelmann}, {Hollows}, {Holmes}, {Holt}, {Holz}, {Hong},
  {Hopkins}, {Hough}, {Hourihane}, {Howell}, {Hoy}, {Hoyland}, {Hreibi},
  {Hsieh}, {Hsu}, {Huang}, {Huang}, {Huang}, {Huang}, {Huang}, {Huang},
  {H{\"u}bner}, {Huddart}, {Hughey}, {Hui}, {Hui}, {Husa}, {Huttner},
  {Huxford}, {Huynh-Dinh}, {Ide}, {Idzkowski}, {Iess}, {Ikenoue}, {Imam},
  {Inayoshi}, {Ingram}, {Inoue}, {Ioka}, {Isi}, {Isleif}, {Ito}, {Itoh},
  {Iyer}, {Izumi}, {JaberianHamedan}, {Jacqmin}, {Jadhav}, {Jadhav}, {James},
  {Jan}, {Jani}, {Janquart}, {Janssens}, {Janthalur}, {Jaranowski}, {Jariwala},
  {Jaume}, {Jenkins}, {Jenner}, {Jeon}, {Jeunon}, {Jia}, {Jin}, {Johns},
  {Johnson-McDaniel}, {Jones}, {Jones}, {Jones}, {Jones}, {Jones}, {Jonker},
  {Ju}, {Jung}, {Jung}, {Junker}, {Juste}, {Kaihotsu}, {Kajita}, {Kakizaki},
  {Kalaghatgi}, {Kalogera}, {Kamai}, {Kamiizumi}, {Kanda}, {Kandhasamy},
  {Kang}, {Kanner}, {Kao}, {Kapadia}, {Kapasi}, {Karat}, {Karathanasis},
  {Karki}, {Kashyap}, {Kasprzack}, {Kastaun}, {Katsanevas}, {Katsavounidis},
  {Katzman}, {Kaur}, {Kawabe}, {Kawaguchi}, {Kawai}, {Kawasaki},
  {K{\'e}f{\'e}lian}, {Keitel}, {Key}, {Khadka}, {Khalili}, {Khan}, {Khazanov},
  {Khetan}, {Khursheed}, {Kijbunchoo}, {Kim}, {Kim}, {Kim}, {Kim}, {Kim},
  {Kim}, {Kimball}, {Kimura}, {Kinley-Hanlon}, {Kirchhoff}, {Kissel}, {Kita},
  {Kitazawa}, {Kleybolte}, {Klimenko}, {Knee}, {Knowles}, {Knyazev}, {Koch},
  {Koekoek}, {Kojima}, {Kokeyama}, {Koley}, {Kolitsidou}, {Kolstein}, {Komori},
  {Kondrashov}, {Kong}, {Kontos}, {Koper}, {Korobko}, {Kotake}, {Kovalam},
  {Kozak}, {Kozakai}, {Kozu}, {Kringel}, {Krishnendu}, {Kr{\'o}lak}, {Kuehn},
  {Kuei}, {Kuijer}, {Kulkarni}, {Kumar}, {Kumar}, {Kumar}, {Kumar}, {Kume},
  {Kuns}, {Kuo}, {Kuo}, {Kuromiya}, {Kuroyanagi}, {Kusayanagi}, {Kuwahara},
  {Kwak}, {Lagabbe}, {Laghi}, {Lalande}, {Lam}, {Lamberts}, {Landry}, {Lane},
  {Lang}, {Lange}, {Lantz}, {La Rosa}, {Lartaux-Vollard}, {Lasky}, {Laxen},
  {Lazzarini}, {Lazzaro}, {Leaci}, {Leavey}, {Lecoeuche}, {Lee}, {Lee}, {Lee},
  {Lee}, {Lee}, {Lee}, {Lehmann}, {Lema{\^\i}tre}, {Leonardi}, {Leroy},
  {Letendre}, {Levesque}, {Levin}, {Leviton}, {Leyde}, {Li}, {Li}, {Li}, {Li},
  {Li}, {Li}, {Lin}, {Lin}, {Lin}, {Lin}, {Lin}, {Linde}, {Linker}, {Linley},
  {Littenberg}, {Liu}, {Liu}, {Liu}, {Liu}, {Llamas}, {Llorens-Monteagudo},
  {Lo}, {Lockwood}, {Loh}, {London}, {Longo}, {Lopez}, {Lopez Portilla},
  {Lorenzini}, {Loriette}, {Lormand}, {Losurdo}, {Lott}, {Lough}, {Lousto},
  {Lovelace}, {Lucaccioni}, {L{\"u}ck}, {Lumaca}, {Lundgren}, {Luo}, {Lynam},
  {Macas}, {MacInnis}, {Macleod}, {MacMillan}, {Macquet}, {Maga{\~n}a
  Hernandez}, {Magazz{\`u}}, {Magee}, {Maggiore}, {Magnozzi}, {Mahesh},
  {Majorana}, {Makarem}, {Maksimovic}, {Maliakal}, {Malik}, {Man}, {Mandic},
  {Mangano}, {Mango}, {Mansell}, {Manske}, {Mantovani}, {Mapelli},
  {Marchesoni}, {Marchio}, {Marion}, {Mark}, {M{\'a}rka}, {M{\'a}rka},
  {Markakis}, {Markosyan}, {Markowitz}, {Maros}, {Marquina}, {Marsat},
  {Martelli}, {Martin}, {Martin}, {Martinez}, {Martinez}, {Martinez},
  {Martinovic}, {Martynov}, {Marx}, {Masalehdan}, {Mason}, {Massera},
  {Masserot}, {Massinger}, {Masso-Reid}, {Mastrogiovanni}, {Matas},
  {Mateu-Lucena}, {Matichard}, {Matiushechkina}, {Mavalvala}, {McCann},
  {McCarthy}, {McClelland}, {McClincy}, {McCormick}, {McCuller}, {McGhee},
  {McGuire}, {McIsaac}, {McIver}, {McRae}, {McWilliams}, {Meacher}, {Mehmet},
  {Mehta}, {Meijer}, {Melatos}, {Melchor}, {Mendell}, {Menendez-Vazquez},
  {Menoni}, {Mercer}, {Mereni}, {Merfeld}, {Merilh}, {Merritt}, {Merzougui},
  {Meshkov}, {Messenger}, {Messick}, {Meyers}, {Meylahn}, {Mhaske}, {Miani},
  {Miao}, {Michaloliakos}, {Michel}, {Michimura}, {Middleton}, {Milano},
  {Miller}, {Miller}, {Miller}, {Millhouse}, {Mills}, {Milotti}, {Minazzoli},
  {Minenkov}, {Mio}, {Mir}, {Miravet-Ten{\'e}s}, {Mishra}, {Mishra}, {Mistry},
  {Mitra}, {Mitrofanov}, {Mitselmakher}, {Mittleman}, {Miyakawa}, {Miyamoto},
  {Miyazaki}, {Miyo}, {Miyoki}, {Mo}, {Modafferi}, {Moguel}, {Mogushi},
  {Mohapatra}, {Mohite}, {Molina}, {Molina-Ruiz}, {Mondin}, {Montani}, {Moore},
  {Moraru}, {Morawski}, {More}, {Moreno}, {Moreno}, {Mori}, {Morisaki},
  {Moriwaki}, {Morr{\'a}s}, {Mours}, {Mow-Lowry}, {Mozzon}, {Muciaccia},
  {Mukherjee}, {Mukherjee}, {Mukherjee}, {Mukherjee}, {Mukherjee}, {Mukund},
  {Mullavey}, {Munch}, {Mu{\~n}iz}, {Murray}, {Musenich}, {Muusse}, {Nadji},
  {Nagano}, {Nagano}, {Nagar}, {Nakamura}, {Nakano}, {Nakano}, {Nakashima},
  {Nakayama}, {Napolano}, {Nardecchia}, {Narikawa}, {Naticchioni}, {Nayak},
  {Nayak}, {Negishi}, {Neil}, {Neilson}, {Nelemans}, {Nelson}, {Nery},
  {Neubauer}, {Neunzert}, {Ng}, {Ng}, {Nguyen}, {Nguyen}, {Nguyen}, {Nguyen
  Quynh}, {Ni}, {Nichols}, {Nishizawa}, {Nissanke}, {Nitoglia}, {Nocera},
  {Norman}, {North}, {Nozaki}, {Nu{\~n}o Siles}, {Nuttall}, {Oberling},
  {O'Brien}, {Obuchi}, {O'Dell}, {Oelker}, {Ogaki}, {Oganesyan}, {Oh}, {Oh},
  {Oh}, {Ohashi}, {Ohishi}, {Ohkawa}, {Ohme}, {Ohta}, {Okada}, {Okutani},
  {Okutomi}, {Olivetto}, {Oohara}, {Ooi}, {Oram}, {O'Reilly}, {Ormiston},
  {Ormsby}, {Ortega}, {O'Shaughnessy}, {O'Shea}, {Oshino}, {Ossokine},
  {Osthelder}, {Otabe}, {Ottaway}, {Overmier}, {Pace}, {Pagano}, {Page},
  {Pagliaroli}, {Pai}, {Pai}, {Palamos}, {Palashov}, {Palomba}, {Pan}, {Pan},
  {Panda}, {Pang}, {Pang}, {Pankow}, {Pannarale}, {Pant}, {Panther},
  {Paoletti}, {Paoli}, {Paolone}, {Parisi}, {Park}, {Park}, {Parker},
  {Pascucci}, {Pasqualetti}, {Passaquieti}, {Passuello}, {Patel}, {Pathak},
  {Patricelli}, {Patron}, {Paul}, {Payne}, {Pedraza}, {Pegoraro}, {Pele},
  {Pe{\~n}a Arellano}, {Penn}, {Perego}, {Pereira}, {Pereira}, {Perez},
  {P{\'e}rigois}, {Perkins}, {Perreca}, {Perri{\`e}s}, {Petermann},
  {Petterson}, {Pfeiffer}, {Pham}, {Phukon}, {Piccinni}, {Pichot},
  {Piendibene}, {Piergiovanni}, {Pierini}, {Pierro}, {Pillant}, {Pillas},
  {Pilo}, {Pinard}, {Pinto}, {Pinto}, {Piotrzkowski}, {Piotrzkowski},
  {Pirello}, {Pitkin}, {Placidi}, {Planas}, {Plastino}, {Pluchar}, {Poggiani},
  {Polini}, {Pong}, {Ponrathnam}, {Popolizio}, {Porter}, {Poulton}, {Powell},
  {Pracchia}, {Pradier}, {Prajapati}, {Prasai}, {Prasanna}, {Pratten},
  {Principe}, {Prodi}, {Prokhorov}, {Prosposito}, {Prudenzi}, {Puecher},
  {Punturo}, {Puosi}, {Puppo}, {P{\"u}rrer}, {Qi}, {Quetschke},
  {Quitzow-James}, {Qutob}, {Raab}, {Raaijmakers}, {Radkins}, {Radulesco},
  {Raffai}, {Rail}, {Raja}, {Rajan}, {Ramirez}, {Ramirez}, {Ramos-Buades},
  {Rana}, {Rapagnani}, {Rapol}, {Ray}, {Raymond}, {Raza}, {Razzano}, {Read},
  {Rees}, {Regimbau}, {Rei}, {Reid}, {Reid}, {Reitze}, {Relton}, {Renzini},
  {Rettegno}, {Reza}, {Rezac}, {Ricci}, {Richards}, {Richardson}, {Richardson},
  {Riemenschneider}, {Riles}, {Rinaldi}, {Rink}, {Rizzo}, {Robertson}, {Robie},
  {Robinet}, {Rocchi}, {Rodriguez}, {Rolland}, {Rollins}, {Romanelli},
  {Romano}, {Romel}, {Romero-Rodr{\'\i}guez}, {Romero-Shaw}, {Romie},
  {Ronchini}, {Rosa}, {Rose}, {Rosi{\'n}ska}, {Ross}, {Rowan}, {Rowlinson},
  {Roy}, {Roy}, {Roy}, {Rozza}, {Ruggi}, {Ruiz-Rocha}, {Ryan}, {Sachdev},
  {Sadecki}, {Sadiq}, {Sago}, {Saito}, {Saito}, {Sakai}, {Sakai},
  {Sakellariadou}, {Sakuno}, {Salafia}, {Salconi}, {Saleem}, {Salemi},
  {Samajdar}, {Sanchez}, {Sanchez}, {Sanchez}, {Sanchis-Gual}, {Sanders},
  {Sanuy}, {Saravanan}, {Sarin}, {Sassolas}, {Satari}, {Sathyaprakash}, {Sato},
  {Sato}, {Sauter}, {Savage}, {Sawada}, {Sawant}, {Sawant}, {Sayah},
  {Schaetzl}, {Scheel}, {Scheuer}, {Schiworski}, {Schmidt}, {Schmidt},
  {Schnabel}, {Schneewind}, {Schofield}, {Sch{\"o}nbeck}, {Schulte}, {Schutz},
  {Schwartz}, {Scott}, {Scott}, {Seglar-Arroyo}, {Sekiguchi}, {Sekiguchi},
  {Sellers}, {Sengupta}, {Sentenac}, {Seo}, {Sequino}, {Sergeev}, {Setyawati},
  {Shaffer}, {Shahriar}, {Shams}, {Shao}, {Sharma}, {Sharma}, {Shawhan},
  {Shcheblanov}, {Shibagaki}, {Shikauchi}, {Shimizu}, {Shimoda}, {Shimode},
  {Shinkai}, {Shishido}, {Shoda}, {Shoemaker}, {Shoemaker}, {ShyamSundar},
  {Sieniawska}, {Sigg}, {Singer}, {Singh}, {Singh}, {Singha}, {Sintes},
  {Sipala}, {Skliris}, {Slagmolen}, {Slaven-Blair}, {Smetana}, {Smith},
  {Smith}, {Soldateschi}, {Somala}, {Somiya}, {Son}, {Soni}, {Soni}, {Sordini},
  {Sorrentino}, {Sorrentino}, {Sotani}, {Soulard}, {Souradeep}, {Sowell},
  {Spagnuolo}, {Spencer}, {Spera}, {Srinivasan}, {Srivastava}, {Srivastava},
  {Staats}, {Stachie}, {Steer}, {Steinhoff}, {Steinlechner}, {Steinlechner},
  {Stevenson}, {Stops}, {Stover}, {Strain}, {Strang}, {Stratta}, {Strunk},
  {Sturani}, {Stuver}, {Sudhagar}, {Sudhir}, {Sugimoto}, {Suh}, {Sullivan},
  {Sullivan}, {Summerscales}, {Sun}, {Sun}, {Sunil}, {Sur}, {Suresh}, {Sutton},
  {Suzuki}, {Suzuki}, {Swinkels}, {Szczepa{\'n}czyk}, {Szewczyk}, {Tacca},
  {Tagoshi}, {Tait}, {Takahashi}, {Takahashi}, {Takamori}, {Takano}, {Takeda},
  {Takeda}, {Talbot}, {Talbot}, {Tanaka}, {Tanaka}, {Tanaka}, {Tanaka},
  {Tanaka}, {Tanasijczuk}, {Tanioka}, {Tanner}, {Tao}, {Tao}, {Tapia San
  Mart{\'\i}n}, {Taranto}, {Tasson}, {Telada}, {Tenorio}, {Terhune},
  {Terkowski}, {Thirugnanasambandam}, {Thomas}, {Thomas}, {Thomas}, {Thompson},
  {Thondapu}, {Thorne}, {Thrane}, {Tiwari}, {Tiwari}, {Tiwari}, {Toivonen},
  {Toland}, {Tolley}, {Tomaru}, {Tomigami}, {Tomura}, {Tonelli},
  {Torres-Forn{\'e}}, {Torrie}, {Tosta e Melo}, {T{\"o}yr{\"a}}, {Trapananti},
  {Travasso}, {Traylor}, {Trevor}, {Tringali}, {Tripathee}, {Troiano},
  {Trovato}, {Trozzo}, {Trudeau}, {Tsai}, {Tsai}, {Tsang}, {Tsang}, {Tsao},
  {Tse}, {Tso}, {Tsubono}, {Tsuchida}, {Tsukada}, {Tsuna}, {Tsutsui},
  {Tsuzuki}, {Turbang}, {Turconi}, {Tuyenbayev}, {Ubhi}, {Uchikata},
  {Uchiyama}, {Udall}, {Ueda}, {Uehara}, {Ueno}, {Ueshima}, {Unnikrishnan},
  {Uraguchi}, {Urban}, {Ushiba}, {Utina}, {Vahlbruch}, {Vajente}, {Vajpeyi},
  {Valdes}, {Valentini}, {Valsan}, {van Bakel}, {van Beuzekom}, {van den
  Brand}, {Van Den Broeck}, {Vander-Hyde}, {van der Schaaf}, {van Heijningen},
  {Vanosky}, {van Putten}, {van Remortel}, {Vardaro}, {Vargas}, {Varma},
  {Vas{\'u}th}, {Vecchio}, {Vedovato}, {Veitch}, {Veitch}, {Venneberg},
  {Venugopalan}, {Verkindt}, {Verma}, {Verma}, {Veske}, {Vetrano},
  {Vicer{\'e}}, {Vidyant}, {Viets}, {Vijaykumar}, {Villa-Ortega}, {Vinet},
  {Virtuoso}, {Vitale}, {Vo}, {Vocca}, {von Reis}, {von Wrangel}, {Vorvick},
  {Vyatchanin}, {Wade}, {Wade}, {Wagner}, {Walet}, {Walker}, {Wallace},
  {Wallace}, {Walsh}, {Wang}, {Wang}, {Wang}, {Ward}, {Warner}, {Was},
  {Washimi}, {Washington}, {Watchi}, {Weaver}, {Webster}, {Weinert},
  {Weinstein}, {Weiss}, {Weller}, {Weller}, {Wellmann}, {Wen}, {We{\ss}els},
  {Wette}, {Whelan}, {White}, {Whiting}, {Whittle}, {Wilken}, {Williams},
  {Williams}, {Williams}, {Williamson}, {Willis}, {Willke}, {Wilson},
  {Winkler}, {Wipf}, {Wlodarczyk}, {Woan}, {Woehler}, {Wofford}, {Wong}, {Wu},
  {Wu}, {Wu}, {Wu}, {Wysocki}, {Xiao}, {Xu}, {Yamada}, {Yamamoto}, {Yamamoto},
  {Yamamoto}, {Yamamoto}, {Yamashita}, {Yamazaki}, {Yang}, {Yang}, {Yang},
  {Yang}, {Yang}, {Yap}, {Yeeles}, {Yelikar}, {Ying}, {Yokogawa}, {Yokoyama},
  {Yokozawa}, {Yoo}, {Yoshioka}, {Yu}, {Yu}, {Yuzurihara}, {Zadro{\.z}ny},
  {Zanolin}, {Zeidler}, {Zelenova}, {Zendri}, {Zevin}, {Zhan}, {Zhang},
  {Zhang}, {Zhang}, {Zhang}, {Zhang}, {Zhao}, {Zhao}, {Zhao}, {Zhao}, {Zheng},
  {Zhou}, {Zhou}, {Zhu}, {Zhu}, {Zimmerman}, {Zlochower}, {Zucker}, \&
  {Zweizig}}]{2021arXiv211103606T}
{The LIGO Scientific Collaboration}, {the Virgo Collaboration}, {the KAGRA
  Collaboration}, {et~al.} 2021, arXiv e-prints, arXiv:2111.03606

\bibitem[{{Valtonen} {et~al.}(2023){Valtonen}, {Zola}, {Gopakumar},
  {L{\"a}hteenm{\"a}ki}, {Tornikoski}, {Dey}, {Gupta}, {Pursimo}, {Knudstrup},
  {Gomez}, {Hudec}, {Jel{\'\i}nek}, {{\v{S}}trobl}, {Berdyugin}, {Ciprini},
  {Reichart}, {Kouprianov}, {Matsumoto}, {Drozdz}, {Mugrauer}, {Sadun},
  {Zejmo}, {Sillanp{\"a}{\"a}}, {Lehto}, {Nilsson}, {Imazawa}, \&
  {Uemura}}]{2023MNRAS.521.6143V}
{Valtonen}, M.~J., {Zola}, S., {Gopakumar}, A., {et~al.} 2023, \mnras, 521,
  6143

\bibitem[{{Wang} {et~al.}(2022){Wang}, {Yin}, {Ma}, \&
  {Wu}}]{2022ApJ...933..225W}
{Wang}, M., {Yin}, J., {Ma}, Y., \& {Wu}, Q. 2022, \apj, 933, 225

\bibitem[{{Wevers} {et~al.}(2022){Wevers}, {Pasham}, {Jalan}, {Rakshit}, \&
  {Arcodia}}]{Wevers2022}
{Wevers}, T., {Pasham}, D.~R., {Jalan}, P., {Rakshit}, S., \& {Arcodia}, R.
  2022, \aap, 659, L2

\bibitem[{{Xian} {et~al.}(2021){Xian}, {Zhang}, {Dou}, {He}, \&
  {Shu}}]{2021ApJ...921L..32X}
{Xian}, J., {Zhang}, F., {Dou}, L., {He}, J., \& {Shu}, X. 2021, \apjl, 921,
  L32

\bibitem[{{Zhao} {et~al.}(2022){Zhao}, {Wang}, {Zou}, {Wang}, \&
  {Dai}}]{2022A&A...661A..55Z}
{Zhao}, Z.~Y., {Wang}, Y.~Y., {Zou}, Y.~C., {Wang}, F. .~Y., \& {Dai}, Z.~G.
  2022, \aap, 661, A55

\end{thebibliography}

\end{document}